\documentclass[aps,prx,twocolumn,reprint,superscriptaddress]{revtex4}
\usepackage{amsmath,amssymb,mathtools,bm,xspace,braket,multirow}
\usepackage{graphicx,hyperref}

\newcommand{\renyi}{R\'enyi\xspace}
\newcommand{\Wg}{\mathrm{Wg}}

\newcommand{\abs}[1]{\left|#1\right|}
\newcommand{\norm}[1]{\left\|#1\right\|}
\newcommand{\tr}[1]{\mathrm{tr}\left[#1\right]}

\newcommand{\Hsp}{\mathcal{H}}

\newcommand{\bop}{\mathcal{B}}
\newcommand{\uop}{\mathcal{U}}
\newcommand{\sop}{\mathcal{A}}
\newcommand{\sPauli}{\mathcal{P}}

\newcommand{\haar}{\mathsf{H}}
\newcommand{\RDU}{\mathsf{D}}
\newcommand{\Pauli}{\mathsf{P}}
\newcommand{\LTE}{\mathsf{L}}
\newcommand{\COE}{\mathrm{COE}}
\newcommand{\CSE}{\mathrm{CSE}}
\newcommand{\ave}[2]{\mathbb{E}_{U\sim#1}\left[#2\right]}

\newcommand{\Hsh}{\mathcal{H}}
\newcommand{\Psh}{P_{\mathrm{sh}}}
\newcommand{\Dsh}{d}

\newtheorem{Def}{Definition}

\begin{document}
\title{Characterizing complexity of many-body quantum dynamics
	\\
	by higher-order  eigenstate thermalization
}

\author{Kazuya Kaneko}
\affiliation{Department of Applied Physics, The University of Tokyo,
	7-3-1 Hongo, Bunkyo-ku, Tokyo 113-8656, Japan
}

\author{Eiki Iyoda}
\affiliation{Department of Physics, Tokai University, \\
	4-1-1 Kitakaname, Hiratsuka-shi, Kanagawa 259-1292, Japan
}

\author{Takahiro Sagawa}
\affiliation{Department of Applied Physics, The University of Tokyo,
	7-3-1 Hongo, Bunkyo-ku, Tokyo 113-8656, Japan
}
	
\date{\today}

\begin{abstract}
Complexity of dynamics is at the core of quantum many-body chaos and exhibits a hierarchical feature: higher-order complexity implies more chaotic dynamics.
Conventional ergodicity in thermalization processes is a manifestation of the lowest order complexity,
which is represented by the eigenstate thermalization hypothesis (ETH)
stating that individual energy eigenstates are thermal.
Here,
we propose a higher-order generalization of the ETH,
named the $ k $-ETH ($ k=1,2,\dots $),
to quantify higher-order complexity of quantum many-body dynamics
at the level of individual energy eigenstates,
where the lowest order ETH (1-ETH) is the conventional ETH.
The explicit condition of  the $k$-ETH is obtained
by comparing Hamiltonian dynamics
with the Haar random unitary of the $k$-fold channel.
As a non-trivial contribution of the higher-order ETH,
we show that
the $ k $-ETH with $ k\geq 2 $
implies a universal behavior of the $ k $th \renyi 
entanglement entropy of individual energy eigenstates.
In particular,
the Page correction of the entanglement entropy
originates from the higher-order ETH,
while as is well known, the volume law can be accounted for by the 1-ETH.
We numerically verify that
the 2-ETH approximately holds for a nonintegrable system,
but does not hold in the integrable case.
To further
investigate the information-theoretic feature behind the $ k $-ETH,
we introduce a concept named a partial unitary $ k $-design
(PU $ k $-design), 
which is an approximation of the Haar random unitary
up to the $ k $th moment, where partial means that only a limited number of observables are accessible.
The $ k $-ETH is a special case of a PU $ k $-design
for the ensemble of Hamiltonian dynamics with random-time sampling.
In addition,
we discuss the relationship between the higher-order ETH and information scrambling
quantified by out-of-time-ordered correlators. 
Our framework provides a unified view on
thermalization,
entanglement entropy,
and unitary $ k $-designs,
leading to deeper characterization of higher-order quantum complexity.

\end{abstract}

\maketitle
\tableofcontents

\section{Introduction
	\label{sec:Introduction}}
Generic nonintegrable quantum many-body systems grow into more and more complex states as time evolves, 
which is the essence of chaos in isolated quantum systems;
a simple and precise characterization of higher-order complexity of many-body Hamiltonian dynamics is desirable for deeper understanding of quantum chaos.
Recently, several fundamental notions have attracted much attention regarding this problem.
One is \textit{quantum ergodicity}
--- the equivalence between the long-time average and the thermal average ---,
which is phrased in terms of the eigenstate thermalization hypothesis (ETH)~\cite{Deutsch1991,Srednicki1994,Rigol2008,DAlessio2015}.
The ETH states that
the expectation value of an observable in an energy eigenstate equals its thermal average,
which is regarded as representing the lowest order complexity of chaotic dynamics.

Another fundamental notion is a \textit{unitary $ k $-design},
which is an ensemble of unitary operators that simulates the Haar random unitary (HRU) up to the $ k $th moment~\cite{DiVincenzo2002,Gross2007,Dankert2009}.
The HRU requires an exponential number of quantum gates to implement,
because it is the most random dynamics with full complexity~\cite{Knill1995},
while unitary $ k $-designs can be approximately implemented by using a polynomial number of quantum gates~\cite{Brandao2016PRL,Brandao2016CMP,Nakata2017}.
It has been argued that unitary $ k $-designs have the fundamental relationship with information scrambling~\cite{Roberts2017,Cotler2017,Hunter-Jones2018,Lloyd2018,Lloyd2018full,Zhuang2019}
quantified by out-of-time-ordered correlators (OTOCs)~\cite{Larkin1969,Kitaev2014,Shenker2014,Roberts2015JHEP,Roberts2015PRL,Hosur2016,Maldacena2016JHEP,Swingle2018},
which is originally motivated by the black hole information paradox~\cite{Hayden2007,Sekino2008,Lashkari2013}.
There are also several studies that discussed the relationship between thermalization and information scrambling in quantum chaotic systems~\cite{Sonner2017,Keyserlingk2018,Nahum2018,Khemani2018,Rakovszky2018,Lewis-Swan2019,Lensky2019,Chan2019,Li2019PRA,Xu2019,Kuo2019}.

\begin{figure}
	\includegraphics[width=1.0\linewidth]{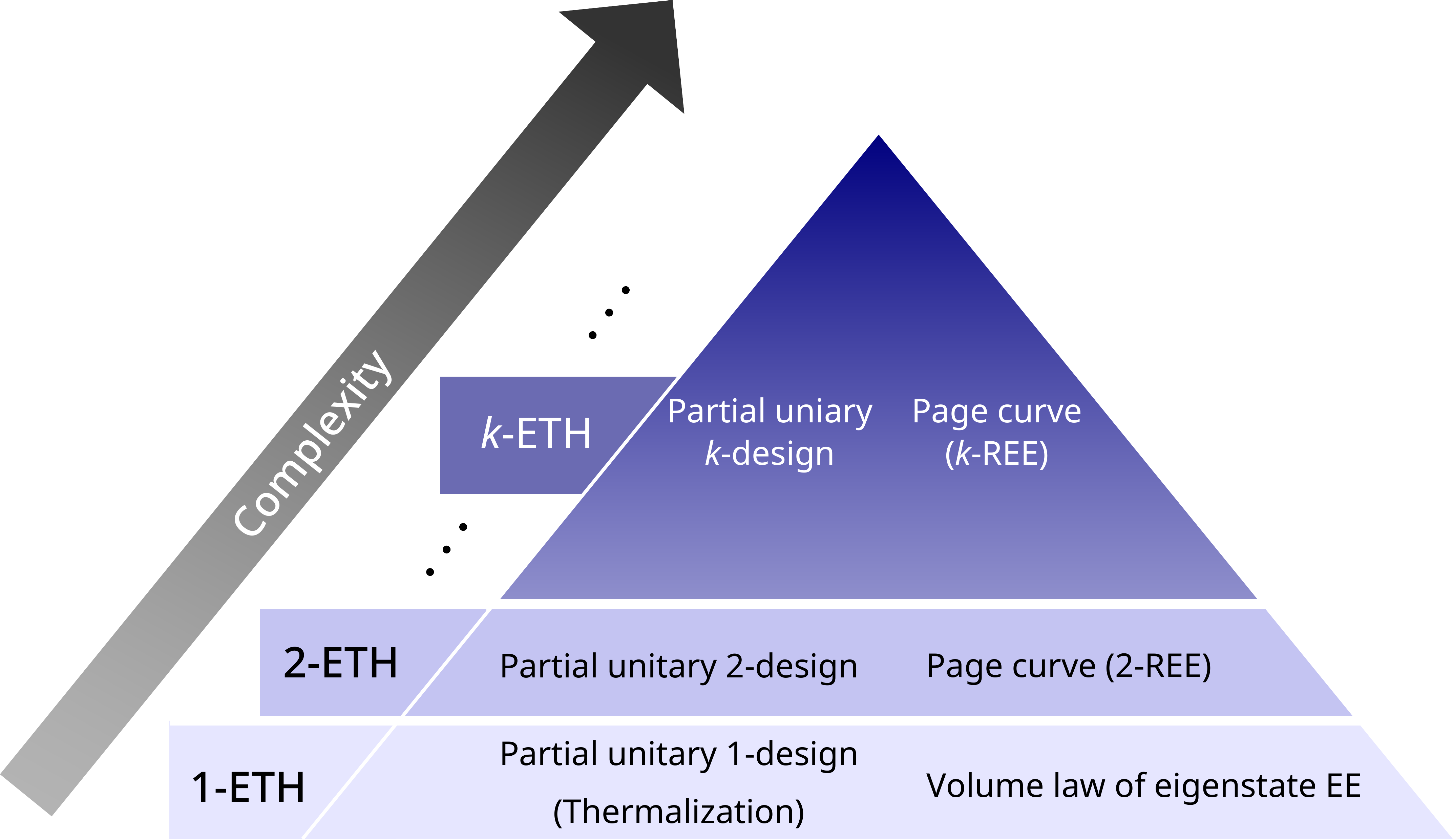}
	\caption{Hierarchy of the $ k $-ETH and PU $ k $-designs.
	The 1-ETH and a PU 1-design reflect the lowest-order complexity of chaotic dynamics,
	i.e., thermalization.
	The 1-ETH also implies that the volume law of the eigenstate entanglement entropy.
	For $ k\geq 2 $, the $ k $-ETH characterizes higher-order complexity related to the Page curve of the $ k $-REE.}
	\label{fig:hierarchy}
\end{figure}

In this paper, we propose a unified framework for the ETH and unitary $ k $-designs by introducing a higher-order generalization of the ETH,
named the $k$-ETH
(see Fig.~\ref{fig:hierarchy}).
The lowest order of the $ k $-ETH, i.e., 1-ETH, is nothing but the conventional ETH.
For $ k\geq 2 $,
the $ k $-ETH represents higher-order complexity than the conventional ergodicity. 
Our formulation is based on the expectation that chaotic dynamics share common properties with the HRU at the level of the higher-order moments described by unitary designs.
We derive the explicit condition of the $k$-ETH by considering a $k$-replicated system and $k$-fold channels
(i.e., a standard technique to deal with the $ k $th moment of dynamics in unitary $ k $-designs).
We point out, however, that
the relevance of the $ k $-ETH to the $ k $th order information scrambling quantified by $ 2k $-point OTOCs~\cite{Roberts2017,Cotler2017} is not straightforward.

We  show that the $ k $-ETH  ($ k\geq 2 $) leads to the universal subsystem-size dependence of the $ k $th \renyi entanglement entropy ($ k $-REE)
of individual energy eigenstates at infinite temperature.
For the $  k$-REE of local regions,
the leading term of the $ k $-ETH leads to the volume law,
which can also be derived from the 1-ETH~\cite{Garrison2018}. 
On the other hand,
for the $ k $-REE of a non-local region
(e.g., the half of a spin chain),
the subleading term of the $ k $-ETH becomes dominant
and gives a non-negligible correction to the volume law,
which corresponds to the Page correction~\cite{Page1993}. 
Thus, we see that the Page correction for a single energy eigenstate is understood
as a unique contribution from the higher-order ETH.
More generally,
the finite-temperature Page curve of the $ k $-REE~\cite{Nakagawa2018,Fujita2018,Lu2019}
is obtained from the $ k $-ETH at finite temperature.

By using numerical exact diagonalization, we verified that the 2-ETH holds for the XXZ ladder model~\cite{Beugeling2014,Beugeling2015,Beugeling2015JSM}.
We found that, in the nonintegrable case, the error of the 2-ETH for a local observable scales as $ \Dsh^{-a} $ with $ a \sim 0.5 $, while as $ d^{-a} $ with $ a\sim 1 $ for a non-local observable, where  $ \Dsh $ is the dimension of the energy shell.
On the other hand, we found that  the 2-ETH does not hold even approximately for an integrable model.
These results imply that the connection between nonintegrability and quantum chaos is true even at a higher-order level than conventional ergodicity.

To look into a more quantum-information theoretic ground of the $k$-ETH, we introduce a \textit{partial} unitary $ k $-design (PU $ k $-design),
which is a slight generalization of a unitary $ k $-design.
It is shown that Hamiltonian dynamics are not unitary designs with random-time sampling,
because the energy conservation does not allow randomizing any energy eigenstate even in the first moment~\cite{Roberts2017}.
However, as long as only a limited number of observables are observed,
it is possible that dynamics can be almost indistinguishable from the HRU,
even if the underlying unitary dynamics itself is not HRU and not even a unitary design.
We formalize such partial imitation of the HRU by a PU $ k $-design.
A conventional unitary $ k $-design is also a PU $ k $-design where all of the observables are observed.
Then, we show that the $k$-ETH is a consequence of a PU $k$-design where the ensemble is given by the random-time sampling of Hamiltonian dynamics.
In particular, from the perspective of a PU $k$-design, we specify the set of observables that satisfy the exact $k$-ETH.

This paper is organized as follows.
In Sec.~\ref{sec:Preliminaries},
we introduce the notation and give the definition of the HRU and the $ k $-fold channel
as a preliminary.
In Sec.~\ref{sec:k-ETH},
we introduce the $ k $-ETH and discuss its relationship to information scrambling.
In Sec.~\ref{sec:Entanglement},
we derive the Page curve of the $ k$-REE
on the basis of the $ k $-ETH.
In Sec.~\ref{sec:Numeric},
we numerically confirm the $ 2 $-ETH
for the nonintegrable model.
In Sec.~\ref{sec:PU-design},
we define PU $ k $-designs and show special examples.
In Sec.~\ref{sec:Discussion},
we summarize our results and discuss future perspectives.
In Appendix~\ref{sec:Proof},
we provide the detail of the proofs of our results.
In Appendix~\ref{sec:Symmetry},
we consider the $ k $-ETH in the presence of
a unitary symmetry as well as an anti-unitary symmetry.
In Appendix~\ref{sec:OSEE},
we show the derivation of the Page curve of
the operator-space entanglement entropy (OSEE) from the $ k $-ETH.
In Appendix~\ref{sec:Supp_numeric}, we provide supplemental numerical results.

\section{Preliminaries
	\label{sec:Preliminaries}}

In this section,
we briefly overview fundamental concepts regarding random unitary dynamics~\cite{Low2010}, and  set our notations used in this paper. 
In particular, we focus on the HRU and the $ k $-fold channel.

Let $ \Hsp $ be a $ d $-dimensional Hilbert space, $ \bop(\Hsp) $ be the set of linear operators on $ \Hsp $,
and $ \uop(\Hsp) $ be the set of unitary operators on $ \Hsp $.
The HRU is the ensemble of unitary operators
sampled according to the Haar measure.
It is defined as the normalized  left and right invariant measure on $ \uop(\Hsp) $:
\begin{align}
\int dU=1,
\
\int f(VU)dU=\int f(UV)dU=\int  f(U)dU,
\label{eq:def_Haar}
\end{align}
for all continuous functions $ f $ and for all $ V\in  \uop(\Hsp)$.
It is known that such a measure is unique.
We denote the ensemble of the HRU by $ \haar $.

\begin{figure}
	\includegraphics[width=0.3\linewidth]{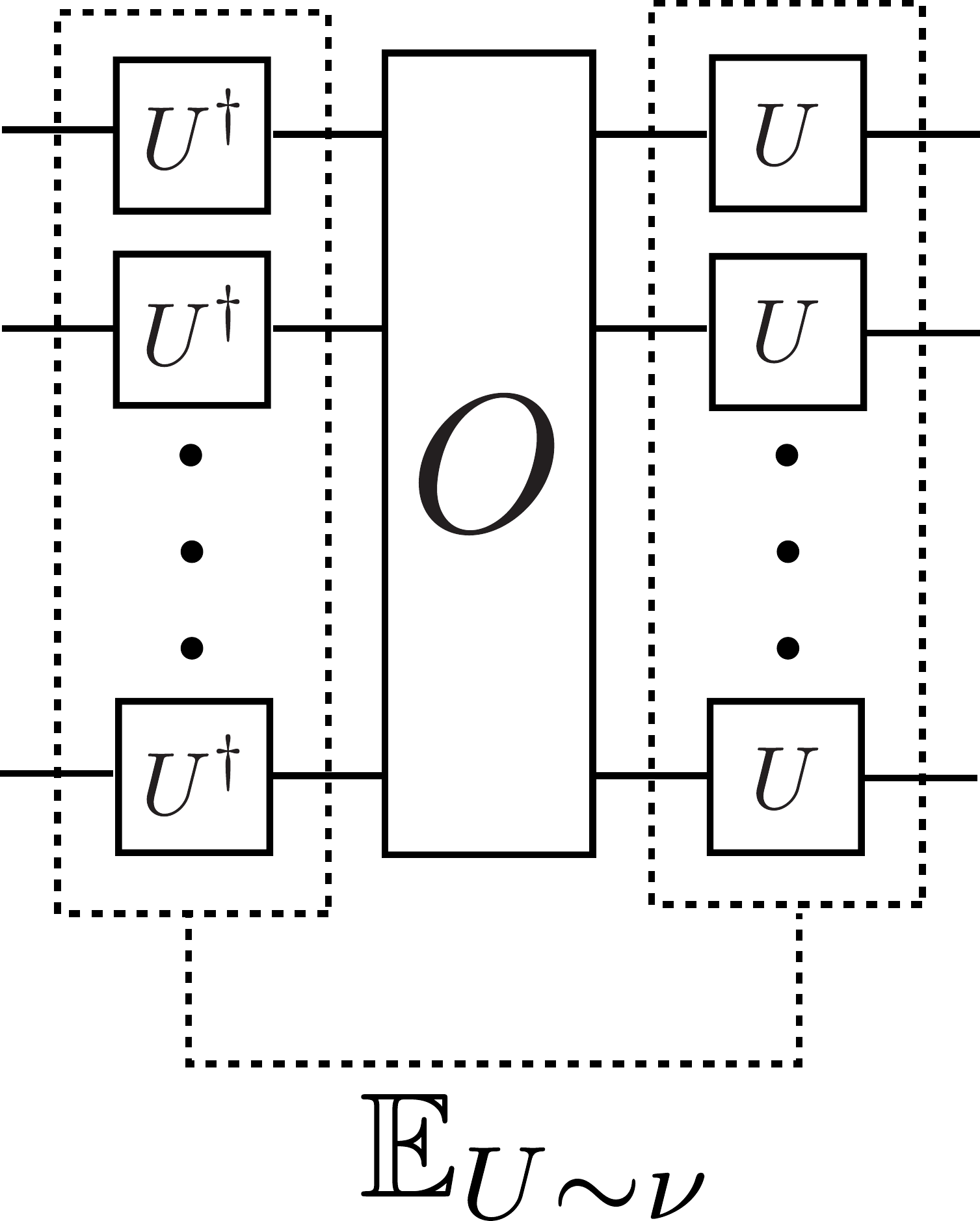}
	\caption{A graphical depiction of the $ k $-fold channel
		\eqref{eq:k-fold_channel}.
	Each horizontal line represents a single Hilbert space $ \Hsp $,
	and ``time'' flies from right to left.}
	\label{fig:k-fold_channel}
\end{figure}	

Let $ \nu $ be an arbitrary ensemble of unitaries sampled from  $ \uop(\Hsp) $.
We denote the ensemble average of $ f(U) $ with $U \in \uop(\Hsp)$ over $ \nu $ by $ \ave{\nu}{f(U)} $.
If $ \nu = \{p_j, U_j\}$ is a discrete ensemble, where unitary $ U_j $ is sampled with probability $ p_j $, the ensemble average is given by
\begin{align}
\ave{\nu}{f(U)} :=\sum_j p_j f(U_j).
\end{align}
If $ \nu = \{p(U), U\}$ is a continuous ensemble,
where $ p(U) $ is a probability density,
then
\begin{align}
\ave{\nu}{f(U)} :=\int dU p(U) f(U).
\end{align}

It is convenient to consider $k$ replicas of $\Hsp$
to treat the $k$th moment of $\nu$.
The $ k $-fold channel $ \Phi^{(k)}_\nu $ with respect to $ \nu $
is defined as
\begin{align}
\Phi^{(k)}_\nu(O)
:=
\ave{\nu}{U^{\dag \otimes k} O U^{\otimes k}},
\label{eq:k-fold_channel}
\end{align}
where $O \in \bop(\Hsp^{\otimes k})$ is an arbitrary operator acting on the $k$ replicas
(see Fig.~\ref{fig:k-fold_channel} for a graphical representation).
The $ k $-fold channel has all the information about
the $ k $th moment of $ \nu $,
and is used to define unitary $ k $-designs
~\cite{DiVincenzo2002,Gross2007,Dankert2009},
as will be discussed in Sec.~\ref{sec:PU-design}.

We next show the explicit form of the $ k $-fold channel of the HRU.
Let $S_{k} $ be the symmetric group of degree $ k $.
We introduce a permutation operator $ W_\pi $
as a unitary representation of a permutation $ \pi\in S_k $,
which acts on $k$ replicas as
\begin{align}
W_\pi \ket{i_1, i_2, \dots, i_k}
=\ket{i_{\pi(1)}, i_{\pi(2)}, \dots, i_{\pi(k)}}
\end{align}
with $ \ket{i_1}, \ket{i_2}, \dots, \ket{i_k}\in \Hsp $.
Due to the unitary invariance \eqref{eq:def_Haar}
of the Haar measure,
$ \Phi_{\haar}^{(k)}(O) $ is invariant under any $ k $-fold unitary conjugation
$ \Phi_{\haar}^{(k)}(O) \mapsto U^{\dag \otimes k} \Phi_{\haar}^{(k)}(O) U^{\otimes k} $.
From the Schur-Weyl duality~\cite{Hayashi2017}, this implies that $ \Phi_{\haar}^{(k)}(O) $ is given by a linear combination of permutation operators.
Thus, we can write the $ k $-fold channel of the HRU as
\begin{align}
\Phi^{(k)}_{\haar}(O)
=\sum_{\pi ,\tau \in S_k}\Wg_{\pi, \tau}(d)\tr{W_\tau O}W_\pi,
\label{eq:Haar_k-channel}
\end{align}
where $ \Wg_{\pi, \tau}(d)$ is the Weingarten matrix~\cite{Colins2006,Gu2013}, which is the inverse of the matrix
whose elements are given by
$ Q_{\pi, \tau}(d) :=\tr{W_\pi W_\tau} $.

The simplest case is the 1-fold channel,
which  is given by
\begin{align}
\Phi^{(1)}_{\haar}(O)=\frac{\tr{O}}{d}I,
\label{eq:1-Haar_channel}
\end{align}
where $ I $ is the identity operator on $ \Hsp $,
and $ \tr{O}/d $ is the expectation value of $ O $
in the maximally mixed state $ I/d $.
The 2-fold channel is given by
\begin{align}
\Phi_{\haar}^{(2)}(O)
&=
\frac{d\tr{O}-\tr{SO}}{d(d^2-1)}I^{\otimes 2}
+\frac{d\tr{SO}-\tr{O}}{d(d^2-1)}S,
\label{eq:2-Haar_channel}
\end{align}
where  $ S $ is the swap operator between the two replicas of $\Hsp$.
As is Fig.~\ref{fig:k-fold_channel},
the swap operator can be graphically 
represented as
\begin{align}
S=	\parbox{0.2\linewidth}{\includegraphics[width=1.0\linewidth]{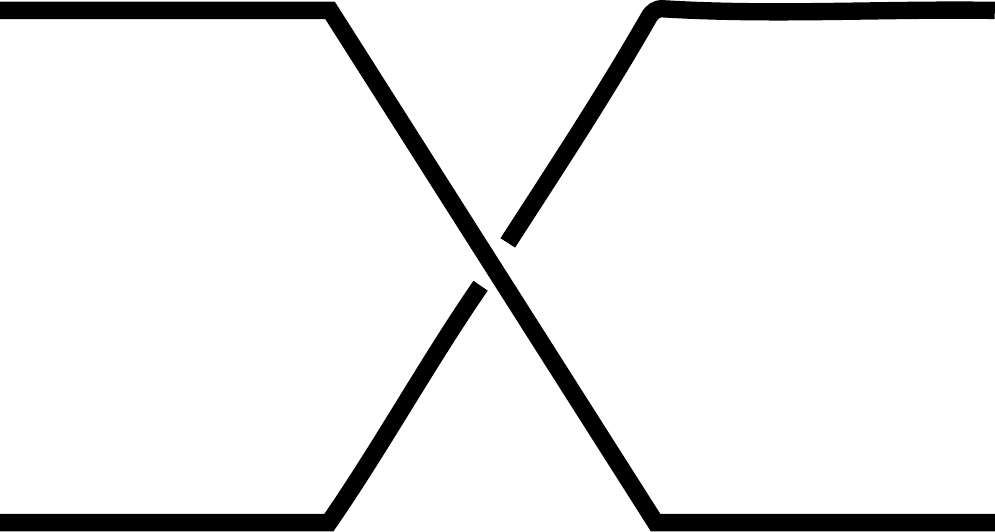}},
\label{eq:swap}
\end{align}
where the upper and lower lines describe the two replica systems.

\section{Higher-order ETH
	\label{sec:k-ETH}}

In this section, we investigate a higher-order generalization of the ETH.
Our basic idea is that chaotic Hamiltonian dynamics can imitate the HRU in the long time run, as long as one observes a few particular observables.
The condition of the $k$-ETH is obtained
by using the $k$-fold channel discussed in Sec.~\ref{sec:Preliminaries}.

\subsection{Long time ensemble}

First of all, we introduce the long-time ensemble (LTE), which is the ensemble of unitaries given by the random-time sampling of Hamiltonian dynamics and describes the late-time behavior of the dynamics.
We consider a quantum many-body system on a lattice with $N$ sites in any spatial dimension.
We assume that there is no symmetry and no local conserved quantities except for the Hamiltonian itself.
We restrict Hamiltonian dynamics  to the energy shell defined below,
reflecting the energy conservation.

Let $ \Hsh $ be a Hilbert space spanned by energy eigenstates
contained in the energy shell
\begin{align}
I_E:=[E-\Delta E, E].
\label{eq:shell}
\end{align}
Here, $ \Delta E $  is the width of the energy shell
and is independent of $ N $.
The dimension of $ \Hsh $ is denoted by $ d $ as before,
which is exponentially large in $ N $.
We assume that, for simplicity, the Hamiltonian has no degeneracy,
and denote an eigenenergy contained in  $ I_E $
by $ E_i\ (i=1,\dots, d) $
and the corresponding eigenstate by $ \ket{E_i} $. 
Then, the Hamiltonian acting on $ \Hsh $ is written as
$ H:= \sum_{i=1}^d E_i\ket{E_i}\bra{E_i}$.

The LTE, denoted as $ \LTE $,
is the ensemble parametrized by time $t$ as
\begin{align}
\LTE
:=
\{e^{-iHt}\}_{t\in (0,\infty)}.
\end{align}
Throughout this paper, we set $ \hbar=1 $.
As seen from the above definition,
the LTE $ \LTE$ describes the uniform sampling of time for Hamiltonian dynamics in the long-time limit.
The ensemble average of a function $f$  with respect to $ \LTE$  is given by
\begin{align}
\ave{\LTE}{f(U)}=\lim_{\tau\to\infty}\frac{1}{\tau}\int_0^{\tau} f(e^{-iHt})dt.
\end{align}

We note that, 
because of the Poincar\'{e}'s recurrence theorem~\cite{Bocchieri1957,Percival1961},
$ e^{-iHt} $ will return to the identity
within an arbitrarily small error at some point.
By considering the LTE, 
however,
we can ignore the effect of shot time intervals in which recurrences occur,
and can characterize the typical behavior of $ e^{-iHt} $ at late times.

A simplest characterization of complexity of the LTE is quantum ergodicity.
Among several definitions of quantum ergodicity~\cite{DAlessio2015,Venuti2019},
we adopt a definition that is directly related to the ETH~\cite{Deutsch1991,Srednicki1994,Rigol2008,DAlessio2015}:
For a given observable $O \in \bop(\Hsh)$, 
the LTE $ \LTE $ satisfies quantum ergodicity,
if the corresponding 1-fold channel satisfies 
\begin{align}
\Phi^{(1)}_\LTE(O)=\Phi^{(1)}_\haar(O),
\label{eq:ergodicity}
\end{align}
where $ \haar $ is the HRU on $ \Hsh $.
The left-hand side above is the long-time average of the Heisenberg evolution of $ O $,
which is given by
\begin{align}
\Phi^{(1)}_\LTE(O)=\lim_{\tau\to\infty}\frac{1}{\tau}\int_0^{\tau} O(t)dt
=\sum_i \ket{E_i}\braket{E_i|O|E_i}\bra{E_i}.
\label{eq:1-LTE_channel}
\end{align}
This definition is similar to that of classical ergodicity, stating that the long-time average equals the microcanonical average.
We note that the validity of Eq.~\eqref{eq:ergodicity}
depends on the choice of  an observable $ O $,
while classical ergodicity is usually regarded as 
an observable-independent concept~\cite{Venuti2019}.

On the other hand, the ETH (strictly speaking, the strong ETH~\cite{Mori2018}) states that for all $ E_i $ in the energy shell~\eqref{eq:shell},
\begin{align}
\braket{E_i|O|E_i}=\braket{O}_{\mathrm{mc}},
\label{eq:ETH}
\end{align}
where $ \braket{O}_{\mathrm{mc}}:= d^{-1}\sum_{i=1}^d\braket{E_i|O|E_i}$
is the microcanonical ensemble average of $ O $.
The ETH~\eqref{eq:ETH} is equivalent to quantum ergodicity in the sense of  Eq.~\eqref{eq:ergodicity},
as seen from Eq.~\eqref{eq:1-Haar_channel} and
Eq.~\eqref{eq:1-LTE_channel}.

\subsection{Definition of $ k $-ETH
\label{sec:k-ETH:definition}}
We now formulate the higher-order extension of quantum ergodicity~\eqref{eq:ergodicity} and the ETH~\eqref{eq:ETH}.
We consider  $k$ replicas of $\Hsh$ and  the $ k $-fold channel~\eqref{eq:k-fold_channel}. 
For a given operator $O \in \bop(\Hsh^{\otimes k})$,
we say that the LTE $ \LTE $ satisfies the $k$th-order quantum ergodicity,
if the corresponding $k$-fold channel satisfies 
\begin{align}
\Phi^{(k)}_\LTE(O)=\Phi^{(k)}_\haar(O).
\label{eq:LTE_k-design}
\end{align}
Equation~\eqref{eq:LTE_k-design}  implies that
the LTE is indistinguishable from the HRU as long as only $ O $ is observed.

To derive a formula for the $k$-ETH, for simplicity, we assume the incommensuration of the energy spectrum~\cite{Srednicki1999,Goldstein2006}:
The spectrum of $ H $ is $ k $th-incommensurate,
if $ \sum_{l=1}^k(E_{i_l}-E_{j_l}) =0$ holds only when
$( i_1,\dots, i_k) $ is a permutation of  $ (j_1,\dots, j_k) $.
This condition is a generalization of the non-resonance condition~\cite{Reimann2008,Linden2009} and ensures the absence of resonances in $ e^{-iHt} $.
The $ k $th-incommensuration is also equivalent to that the LTE is indistinguishable from the ensemble of random unitaries that are diagonal in the energy eigenbasis,
given by $ U=\sum_je^{i\phi_j}\ket{E_j}\bra{E_j} $ with $  \phi_j$ distributing uniformly over $ [0,2\pi] $, up to the $ k $th moment,
(i.e., a diagonal-unitary $ k $-design~\cite{Nakata2017,Nakata2012,Nakata2014}).
We can expect that generic nonintegrable systems have incommensurate spectrum,
while integrable systems do not in general~\cite{Srednicki1999,Goldstein2006}.

Under the $ k $th incommensurate condition,
we obtain the following necessary and sufficient condition for Eq.~\eqref{eq:LTE_k-design} :
for any $ i_1,\dots, i_k $
and $ \sigma\in S_k $,
\begin{align}
&\braket{E_{i_1}\dots E_{i_k}|O|E_{i_{\sigma(1)}}\dots E_{i_{\sigma(k)}}}
\nonumber
\\
&\qquad=
\sum_{\pi,\tau \in S_k}
\delta_{\pi\sigma}(\bm{i})\Wg_{\pi,\tau}(d)\tr{W_{\tau}O},
\label{eq:k-ETH}
\end{align}
where
$ \delta_{\sigma}(\bm{i}):=\delta_{i_1i_{\sigma(1)}}\dots \delta_{i_ki_{\sigma(k)}} $.
The proof is presented in Appendix~\ref{sec:Proof:k-ETH}.
We call Eq.~\eqref{eq:k-ETH} the $ k $-ETH for $ O $.
For $ k=1 $,
we recover the conventional ETH
\eqref{eq:ETH},
which can be referred to as the 1-ETH.
Although the $ 1 $-ETH concerns only the diagonal elements of $ O $,
the $ k $-ETH with $ k\geq 2 $ gives a condition about the off-diagonal elements as well, which will be explicitly shown for $k=2$ later.
We note that, we need to project $ O $ in right-hand side of Eq.~\eqref{eq:k-ETH} to the Hilbert space $ \Hsh^{\otimes k} $ of the energy shell~\eqref{eq:shell}.

As will be discussed in Sec.~\ref{sec:Entanglement} in detail,
the $ k $-ETH ($k \geq 2$) gives a non-negligible contribution in the case of non-local operators acting among multiple replicas. 
For example, we will show that the $k$-ETH of the partial swap operator leads to the Page correction of the $k$-REE.
In such a case, the $ k $-ETH cannot reduce to the conventional ETH, which is the situation that we mainly focus on in this paper.
On the other hand, if the operator is given by a tensor product of the replicas, i.e., $O = A^{\otimes k}$ with $ A\in \bop(\Hsh)$,
the $ k $-ETH reduces to the 1-ETH and the off-diagonal ETH~\cite{Srednicki1994,DAlessio2015} up to the subleading correction, as shown below.

Let $ O=A^{\otimes k}$.
Then, the $ k$-ETH~\eqref{eq:k-ETH} is rewritten as
\begin{align}
&A_{i_1i_{\sigma(1)}}\cdots A_{i_ki_{\sigma(k)}}
\nonumber\\
&\qquad=
\sum_{\pi,\tau \in S_k}
\delta_{\pi\sigma}(\bm{i})\Wg_{\pi,\tau}(d)
\prod_{m=1}^{c(\tau)}\tr{A^{l_m(\tau)}},
\label{eq:k-ETH_product}
\end{align}
where $ A_{ij}:=\braket{E_i|A|E_j} $, $ c(\tau) $ is the number of cycles in the representation of $ \tau $ of the products of cycles,
and $ l_m(\tau) $ is the length of the $ m $th cycle.
We now consider the asymptotic behavior of the right-hand side of Eq.~\eqref{eq:k-ETH_product} with $ \Dsh\to \infty $, where we use the Landau symbol $ \mathcal{O}(\cdot) $.
Assuming that $ \braket{A^l}_{\mathrm{mc}}:=\tr{A^l}/\Dsh=\mathcal{O}(1) $ for any $ l=1,\dots, k $,
we have
\begin{align}
&A_{i_1i_{\sigma(1)}}\cdots A_{i_ki_{\sigma(k)}}
\nonumber\\
&\quad=
\begin{dcases}
\braket{A}_{\mathrm{mc}}^{k}+\mathcal{O}(\Dsh^{-1})
\ & \text{for}\ i_1=i_{\sigma(1)},\dots, i_k=i_{\sigma(k)}
\\
\mathcal{O}(\Dsh^{-1})
& \text{otherwise}.
\end{dcases}
\label{eq:k-ETH_product_asymptotic}
\end{align}
The proof of this is again given in Appendix~\ref{sec:Proof:product}.
For $ k\geq 2 $, this result implies that $ A_{ii}\to \braket{A}_{\mathrm{mc}} $ and $ A_{ij}=\mathcal{O}(\Dsh^{-1/2})\  (i\neq j) $ in the limit of $ \Dsh\to\infty $.
The former is equivalent to the 1-ETH~\eqref{eq:ETH}, and the latter is the scaling consistent with the off-diagonal ETH~\cite{Srednicki1994,DAlessio2015}.
In this sense, the $ k $-ETH for  $ A^{\otimes k} $ reduces to the conventional ETH.

We remark that
a specific case of  the higher-order extension of the ETH is also discussed in Ref.~\cite{Foini2018}.
They derived the higher-order ETH for the product of $ k $  matrix elements with cyclic indices, i.e., $ A_{i_1i_2}A_{i_2i_3}\dots A_{i_ki_1} $.
This is a specific case of Eq.~\eqref{eq:k-ETH_product} where $ \sigma $ is a cyclic permutation, which reduces to the conventional ETH as discussed above.
On the other hand, our proposal \eqref{eq:k-ETH} is more general and applicable to the operators which cannot be represented as a tensor product  of operators acting on the individual Hilbert spaces.

As is the case for the 1-ETH, the $ k $-ETH \eqref{eq:k-ETH} does not exactly hold for generic many-body but finite-size systems.
Thus, an important point is whether Eq.~\eqref{eq:k-ETH} asymptotically holds in the limit of $ \Dsh\to\infty $, which brings us  to investigate the finite-size scaling of the error in  Eq.~\eqref{eq:k-ETH},
as will be discussed  in Sec.~\ref{sec:k-ETH:Approximate_k-ETH}.

We note that 
the $ k $-ETH differs from the 1-ETH of $k$-replicas with Hamiltonian
$ H_k:=H\otimes I^{\otimes (k-1)}+\cdots  +I^{\otimes (k-1)}\otimes H $.
In fact, the corresponding 1-ETH is given by
\begin{align}
\braket{E_{i_1}\dots E_{i_k}|O|E_{i_1}\dots E_{i_k}}
=
\frac{\tr{O}}{\Dsh^k}.
\label{eq:non-interacting_k-ETH}
\end{align}
The right-hand-side of Eq.~\eqref{eq:non-interacting_k-ETH} is the average of $ O $ in the maximally mixed state of $ \Hsh^{\otimes k} $,
which does not equal the right-hand side of Eq.~\eqref{eq:k-ETH}.
If $ \ket{E_{i_1}},\dots, \ket{E_{i_k}} $ were independent and identically distributed (i.i.d.) random states,
Eq.~\eqref{eq:non-interacting_k-ETH} would be true.
In reality, however, energy eigenstates are not independent due to their orthogonality condition.
This fact does not affect the first moment (and thus the 1-ETH),
while cannot be ignored for the higher moments (and thus the higher-order ETH).

We finally note that
Eq.~\eqref{eq:LTE_k-design} is very similar to the definition of a unitary $ k $-design,
which requires that Eq.~\eqref{eq:LTE_k-design} holds for all operators.
Indeed, we discuss the relationship between the $ k $-ETH and unitary $ k $-designs in Sec.~\ref{sec:PU-design}.

\subsection{Approximate $ k$-ETH
\label{sec:k-ETH:Approximate_k-ETH}}
 
As already mentioned,
the $ k $-ETH does not exactly hold in general when $ d $ is finite.
Thus, the validity of the $ k $-ETH should be investigated through the scaling of the  system-size dependence.
We can say that the $ k $-ETH approximately holds when the error of the $ k $-ETH vanishes in $ d\to\infty $.
In the following,
we introduce the indicator to evaluate the error of the $ k $-ETH in a similar fashion to the 1-ETH.

For the 1-ETH, several indicators of the ETH have been proposed for numerical studies~\cite{Steinigeweg2013,Beugeling2014,Kim2014,Ikeda2015,Mondaini2016,Dymarsky2018,Yoshizawa2018}.
Among them, we here focus on the indicator~\cite{Ikeda2015}
\begin{align}
I_1(O):=\max_{i\in\{1,\dots, \Dsh\}}
\abs{\braket{E_i|O|E_i}-\braket{O}_{\mathrm{mc}}}.
\end{align}
The 1-ETH is true in the strong sense (i.e., there does not exist any non-thermal energy eigenstate), if and only if  $ I_1(O )$ vanishes with $ \Dsh\to\infty $.
Here, we used the assumption that the width of the energy shell $\Delta E$ is $\mathcal O(1)$.
We note that $ I_1(O) $ is the strongest  indicator of the ETH, because $ I_1(O) \to 0$ implies the decay of the other indicators~\cite{Mori2018}.
In particular,
it has been numerically confirmed that  $ I_1(O )$ decays in generic nonintegrable systems in the absence of any exceptional ``scar'' state~\cite{Heller1984,Bernien2017,Turner2018},
while does not decay in integrable systems~\cite{Ikeda2015}.

We now introduce an indicator of the $ k $-ETH as a natural generalization of $ I_1(O )$.
For $ O\in \bop(\Hsh^{\otimes k}) $, we define
\begin{align}
I_k(O):=
\max_{\substack{i_1,\dots, i_k\in\{1,\dots, \Dsh\}\\ \sigma\in S_k}}
\abs{\Delta(i_1,\dots, i_k;\sigma)},
\label{eq:k-ETH_indicator}
\end{align}
where $ \Delta(i_1,\dots, i_k;\sigma) $ is the difference between the both-hand sides of
Eq.~\eqref{eq:k-ETH}, i.e.,
\begin{align}
\Delta(i_1,\dots, i_k;\sigma)
=
&\braket{E_{i_1}\dots E_{i_k}|O|E_{i_{\sigma(1)}}\dots E_{i_{\sigma(k)}}}
\nonumber\\
&\qquad-\sum_{\pi,\tau \in S_k}\delta_{\pi\sigma}(\bm{i})\Wg_{\pi,\tau}(d)\tr{W_{\tau}O}.
\label{eq:k-ETH_error}
\end{align}
We note that $ I_k(O) $ equals the maximum norm of operator $\Delta^{(k)}(O):= \Phi^{(k)}_\LTE(O)-\Phi^{(k)}_\haar(O) $.
The $ k $-ETH~\eqref{eq:k-ETH} exactly holds if and only if $ I_k(O)=0 $.
The validity of the $ k $-ETH for many-body systems can be judged by looking at weather $ I_k(O) $ converges to 0 with $ \Dsh\to\infty $.
Thus, we say that the $k$-ETH is \textit{approximately} true, if $\lim_{\Dsh \to \infty} I_k (O) = 0$.
On the other hand, we say that the \textit{exact} $ k $-ETH is true if $ I_k(O)=0 $.

We note that the approximate $ k $-ETH is typically true for random operators.
To see this, we consider the uniformly random sampling of operator $O$ from the normalized operator space of $  \bop(\Hsh^{\otimes k}) $ with respect to the Hilbert-Schmidt norm $\norm{O}^2_{\mathrm{HS}}:=\Dsh^{-k} \mathrm{tr}[O^\dag O] $.
Then, we can show that
\begin{align}
I_k(O) =\mathcal{O}(\Dsh^{-k/2+\delta})
\label{eq:typical_error}
\end{align}
holds with a probability close to 1, where $ \delta>0 $ is an arbitrary positive constant (see Appendix~\ref{sec:Proof:typical_error} for the proof).
For $ k=1 $, Eq.~\eqref{eq:typical_error} recovers the known result for the 1-ETH, $ I_1(O) =\mathcal{O}(\Dsh^{-1/2+\delta}) $~\cite{Reimann2015}.

To obtain the above scaling~\eqref{eq:typical_error}, the operator $O$ is sampled from the operator space on $\Hsh^{\otimes k}$. 
Thus, we expect that the scaling~\eqref{eq:typical_error} can be confirmed in nonintegrable systems, by choosing $O$ as a generic operator that is not in the tensor product form $O = A^{\otimes k}$ and collectively acts on the $k$ replicas. 
Indeed, we will numerically show the scaling of Eq.~\eqref{eq:typical_error}  with $k=2$ for a non-local partial swap operator in Sec.~\ref{sec:Numeric}.

On the other hand, if the operator is given by  the tensor product form $O = A^{\otimes k}$, it does not correspond to a typical operator on $  \bop(\Hsh^{\otimes k}) $, and thus exhibits a different scaling from Eq.~\eqref{eq:typical_error}.
In fact, random matrix theory for the Hamiltonian predicts that $ \abs{A_{ii}-\braket{A}_{\mathrm{mc}}}=\mathcal{O}(\Dsh^{-1/2})$ and $ \abs{A_{ij}}=\mathcal{O}(\Dsh^{-1/2})\ (i\neq j)$~\cite{DAlessio2015}.
Thus, from  Eq.~\eqref{eq:k-ETH_product_asymptotic}, $ |A_{ii}^k-\braket{A}_{\mathrm{mc}}^k|=\mathcal{O}(\Dsh^{-1/2}) $  leads to $ I_k(A^{\otimes k})= \mathcal{O}(\Dsh^{-1/2})$ for $ \abs{\braket{A}_{\mathrm{mc}}}=\mathcal{O}(1) $, and $ \abs{A_{ii}}=\mathcal{O}(\Dsh^{-1/2}) $ leads to  $ I_k(A^{\otimes k})= \mathcal{O}(\Dsh^{-1})$ for $ \braket{A}_{\mathrm{mc}}=0 $.
If $ \abs{\braket{A}_{\mathrm{mc}}}=\mathcal{O}(1) $ and $ k\geq 2 $, this scaling is different from that obtained for random operators~\eqref{eq:typical_error}, because $ A^{\otimes k} $ is not a typical operator in $  \bop(\Hsh^{\otimes k}) $ as mentioned above.
In fact, the measure of operators given by the tensor product form in $  \bop(\Hsh^{\otimes k}) $  is zero.
More generally, if $ O $ is a linear combination of $ \mathcal{O}(1) $ operators in the form of $ A^{\otimes k} $, we have the above scaling.

In addition,  we consider the case that the microcanonical average itself scales as $ \abs{\braket{A}_{\mathrm{mc}}}=\mathcal{O}(\Dsh^{-\alpha}) $ with some constant $ 0\leq \alpha< 1/2 $.
As an example, let us consider the microcanonical average of a local Pauli operator at infinite temperature. 
Its average is exactly zero if we take the average over the completely mixed state of the entire Hilbert space (i.e., the microcanonical average including negative temperature states).
On the other hand, if we take the microcanonical average with the energy shell~\eqref{eq:shell}, the average can deviate from zero because of the finite-size effect.
These averages coincide in the thermodynamic limit because of the equivalence of ensembles~\cite{Ruelle1999,Tasaki2018}, and thus we can expect the above-mentioned scaling.
In this case, we have $ |A_{ii}^k-\braket{A}_{\mathrm{mc}}^k|=\mathcal{O}(\Dsh^{-1/2-\alpha(k-1)})$,
and thus
\begin{align}
I_k(A^{\otimes k})=\mathcal{O}(\Dsh^{-1/2-\alpha(k-1)}).
\label{eq:scaling_RMT_product}
\end{align}

\subsection{2-ETH
\label{sec:k-ETH:2-ETH}}
As a special case, we consider the $2 $-ETH in detail.
For $ k=2 $, the general condition~\eqref{eq:k-ETH} reduces to 
\begin{align}
\braket{E_iE_i|O|E_iE_i}
&=
\frac{\tr{O}+\tr{SO}}{\Dsh(\Dsh+1)},
\label{eq:2-ETH_1}
\\
\braket{E_iE_j|O|E_iE_j}
&=
\frac{\Dsh\tr{O}-\tr{SO}}{\Dsh(\Dsh^2-1)}
\quad (i\neq j),
\label{eq:2-ETH_2}
\\
\braket{E_iE_j|O|E_jE_i}
&
=
\frac{\Dsh\tr{SO}-\tr{O}}{\Dsh(\Dsh^2-1)}
\quad (i\neq j)
\label{eq:2-ETH_3},
\end{align}
where $O \in \bop(\Hsh^{\otimes 2})$, and $ S $ is the swap operator~\eqref{eq:swap}.
In the above condition, there are totally $ d^2 $ equalities for the diagonal elements of $ O $ (Eqs.~\eqref{eq:2-ETH_1} and \eqref{eq:2-ETH_2}) and  $ d^2-d $ equalities for the off-diagonal elements of $ O $ (Eq.~\eqref{eq:2-ETH_3}).

As a specific (and not generic) case, let  $ O=A^{\otimes 2}$ with $ A\in \bop(\Hsh)$.
In this case, 
Eqs. \eqref{eq:2-ETH_1}-\eqref{eq:2-ETH_3} reduce to 
\begin{align}
A_{ii}A_{ii}
&=
\braket{A}_{\mathrm{mc}}^2
+
\frac{\braket{A^2}_{\mathrm{mc}}-\braket{A}_{\mathrm{mc}}^2}{\Dsh+1},
\label{eq:2-ETH_1_product_op}
\\
A_{ii}A_{jj}
&=
\braket{A}_{\mathrm{mc}}^2
-
\frac{\braket{A^2}_{\mathrm{mc}}-\braket{A}_{\mathrm{mc}}^2}{\Dsh^2-1}
\quad (i\neq j),
\label{eq:2-ETH_2_product_op}
\\
A_{ij}A_{ji}
&=
\frac{d(\braket{A^2}_{\mathrm{mc}}-\braket{A}_{\mathrm{mc}}^2)}{d^2-1}
\quad (i\neq j)
\label{eq:2-ETH_3_product_op},
\end{align}
where we used 
$ \tr{SA^{\otimes 2}}=\tr{A^2}=d\braket{A^2}_{\mathrm{mc}} $.
We note that these equalities has been obtained from random matrix theory in Ref.~\cite{Hamazaki2019}.
We remark that
the 2-ETH for $ A^{\otimes 2} $
is not exactly compatible with the 1-ETH for $ A $.
In fact, the 1-ETH for $ A $ gives $A_{ii}A_{jj}=\braket{A}_{\mathrm{mc}}^2 $ for all $ i, j $.
This deviates from the 2-ETH by the second terms on the right-hand sides of Eqs.~\eqref{eq:2-ETH_1_product_op} and \eqref{eq:2-ETH_2_product_op},
which are not zero in general.
With $ \Dsh\to\infty $, however, these terms vanish if $ \braket{A}_{\mathrm{mc}}=\mathcal{O}(1) $ and $ \braket{A^2}_{\mathrm{mc}}=\mathcal{O}(1) $.
Therefore,
the 2-ETH is compatible with the 1-ETH only approximately.
In other words, the 2-ETH for $ A^{\otimes 2} $ only gives the subleading correction of the order $ \mathcal{O}(d^{-1}) $ to the 1-ETH.

We next discuss the approximate 2-ETH as a special case of the general argument in Sec.~\ref{sec:k-ETH:Approximate_k-ETH}.
The indicator~\eqref{eq:k-ETH_indicator} now reduces to
\begin{align}
I_2(O)
=\max_{i,j\in\{1,\dots, \Dsh\}}
\left[\abs{\Delta_i^{1}},\abs{\Delta_{ij}^{2}},\abs{\Delta_{ij}^{3}}\right],
\label{eq:2-ETH_indicator}
\end{align}
where
\begin{align}
\Delta_i^{1}
&:=
\braket{E_iE_i|O|E_iE_i}-\frac{\tr{O}+\tr{SO}}{\Dsh(\Dsh+1)},
\label{eq:2-ETH_1_error}
\\
\Delta_{ij}^{2}
&:=
\braket{E_iE_j|O|E_iE_j}-\frac{\Dsh\tr{O}-\tr{SO}}{\Dsh(\Dsh^2-1)},
\label{eq:2-ETH_2_error}
\\
\Delta_{ij}^{3}
&:=
\braket{E_iE_j|O|E_jE_i}-\frac{\Dsh\tr{SO}-\tr{O}}{\Dsh(\Dsh^2-1)}.
\label{eq:2-ETH_3_error}
\end{align}
In our numerical calculation by exact diagonalization (presented in Sec.~\ref{sec:Numeric}), we find that $ I_2(O )$ decays polynomially with $ \Dsh $ for several operators of a nonintegrable system and does not decay in an integrable system.

\subsection{$ k $-ETH and information scrambling
\label{sec:k-ETH:scrambling}}
We discuss the relationship between the $ k $-ETH
and information scrambling.
Considering the $ 2k $-point OTOC as 
an indicator of  higher-order scrambling~\cite{Roberts2017,Cotler2017},
we show that the exact $ k $-ETH is a sufficient condition for
the decay of the  $ 2k $-point OTOC to the exact value of the HRU average.
However,
we point out that
the approximate decay of the $ 2k $-point OTOC follows 
only from the conventional diagonal and  off-diagonal ETH,
and the unique role of the higher-order ETH is smaller than the finite-size effect as for the OTOC.

We define the $ 2k $-point OTOC~\cite{Roberts2017,Cotler2017} for $ A, B \in\bop(\Hsh) $ by
\begin{align}
F_{A,B}^{(2k)}(t):=
\braket{A(t)B\cdots A(t)B}_{\mathrm{mc}},
\label{eq:2k-OTOC}
\end{align}
where the most commonly discussed case is $ k=2 $.
We also consider its HRU average defined as
\begin{align}
F_{A,B}^{(2k),\haar}:=
\ave{\nu}{\braket{(UAU^\dag)B\cdots
	(UAU^\dag)B}_{\mathrm{mc}}}.
\label{eq:OTOC_HRU}
\end{align}
If the $ 2k $-point OTOC decays to the exact HRU average,
we say that dynamics exhibits the exact $ k $th-order information scrambling. 
It is known that
if the dynamics is a unitary $ k $-design,
the $ 2k $-point OTOC equals the exact HRU average value~\cite{Roberts2017,Cotler2017}.

We consider the long-time average of  the 2$ k $-point OTOC
\begin{align}
\overline{F}_{A,B}^{(2k)}:=
\lim_{T\to\infty}\frac{1}{T}\int_0^T F_{A,B}^{(2k)}(t)dt,
\label{eq:OTOC_LTA}
\end{align}
which characterizes the typical late-time behavior of $ F_{A,B}^{(2k)}(t) $.
As represented by Eq.~\eqref{eq:LTE_k-design},
the $k$-ETH implies that the LTE is indistinguishable from the HRU
if the $k$-fold channel is applied.
This suggests that the $k$-ETH also implies that
the long-time average of the $2k$-point OTOC equals its HRU average.
In fact,
if the exact $ k $-ETH~\eqref{eq:k-ETH} for $ O=A^{\otimes k} $ is satisfied,
we have
\begin{align}
\overline{F}_{A,B}^{(2k)}
=
F_{A,B}^{(2k),\haar}
\label{eq:exact_scrambling}
\end{align}
for any $ B\in\bop(\Hsh) $ (see Appendix~\ref{sec:Proof:exact_scrambling} for the proof).
This implies that the exact $ k $-ETH is a sufficient condition for the exact $ k $th-order information scrambling at late times,
which is a generalization of  the fact that the 1-ETH is a sufficient condition for thermalization.
In other words, the $ k $-ETH represents the $ k $th-order information scrambling at the level of individual energy eigenstates.

As mentioned in Sec.~\ref{sec:k-ETH:Approximate_k-ETH},
however,
the exact $k$-ETH is hardly satisfied for many-body quantum systems.
It is thus crucial to ask whether Eq.~\eqref{eq:exact_scrambling} is approximately satisfied if the system size is large.
For $ k=2 $,
Ref.~\cite{Huang2019PRL} has provided an affirmative answer only by assuming the approximate 1-ETH:
If the approximate $1$-ETH for $ A, A^2, B, B^2 $ is true,
then
\begin{align}
\overline{F}_{A,B}^{(4)}
=
F_{A,B}^{(4),\haar}
+o(1),
\end{align}
where $ o(1) $ denotes a function that vanishes with $ \Dsh\to\infty $.
In terms of the $ k $-ETH, their result is understood from the fact that the approximate $ 2 $-ETH for $ A^{\otimes k} $ is reduced to the approximate 1-ETH as shown in Sec.~\ref{sec:k-ETH:Approximate_k-ETH}.

For $ k\geq 3 $ with traceless operators $ A, B \in\bop(\Hsh) $,
we can show the following statement (see Appendix~\ref{sec:Proof:approximate_scrambling} for the proof).
Suppose that, for $O= A , B$, the 1-ETH is approximately true with scaling of the form $ \abs{O_{ii}-\braket{O}_{\mathrm{mc}}}=\mathcal{O}(\Dsh^{-1/2}) $
and that the off-diagonal ETH with $ \abs{O_{ij}}=\mathcal{O}(\Dsh^{-1/2}) $ for $ i\neq j $ is true as in Sec.~\ref{sec:k-ETH:Approximate_k-ETH}.
Then, we can show that
\begin{align}
\overline{F}_{A,B}^{(2k)}
= \mathcal{O}(\Dsh^{-1}),
\label{eq:approximate_scrambling}
\end{align}
which is slower than the decay of the HRU average
$ F_{A,B}^{(2k),\haar}= \mathcal{O}(\Dsh^{-2})$~\cite{Roberts2017}.

The foregoing results  imply that for any $k$, the approximate decay of the $2k$-point OTOC is a consequence of the conventional ETH rather than the higher-order ETH, while the exact decay requires the exact $ k $-ETH.
Since the $ k $-ETH holds only approximately in generic many-body systems, the exact decay to the HRU average would never be achieved.
Moreover, as shown in Appendix~\ref{sec:Supp_numeric}, we numerically confirmed that the decay of the 4-point OTOC in a nonintegrable many-body system is consistent with the prediction from the conventional ETH~\eqref{eq:approximate_scrambling}.
Therefore, we conclude that the relationship between the $k$th-order information scrambling and the $k$-ETH (and thus unitary $k$-designs) is not straightforward.

We note that information scrambling is also characterized by the operator-averaged OTOC~\cite{Hosur2016,Yoshida2017}, which is equivalent to the operator space entanglement entropy (OSEE) of unitary operators~\cite{Zanardi2001,Prosen2007}.
In Appendix~\ref{sec:OSEE}, 
we will prove that the page curve of the $ k $th \renyi OSEE ($ k $-OSEE) follows from the $ k $-ETH.

\section{Eigenstate entanglement
	\label{sec:Entanglement}}

In this section,
we derive the Page curve of the $ k $-REE ($ k\geq 2 $) in individual energy eigenstates at infinite temperature, on the basis of the $ k $-ETH for the partial cyclic permutation operator.
The Page curve has been originally proposed as the average entanglement entropy of Haar random states~\cite{Page1993}, while we here focus on the entanglement entropy of a single energy eigenstate as numerically studied in Refs.~\cite{Garrison2018,Beugeling2015JSM,Vidmar2017,Vidmar2017quad,Vidmar2018,Hackl2019,LeBlond2019}.
In addition, we consider the case of finite-temperature eigenstates and derive a generalized Page curve~\cite{Nakagawa2018,Fujita2018,Lu2019,Huang2019,Murthy2019PRE}.

We start with the definition of the partial cyclic permutation operator and the $ k $-REE.
For simplicity, let us consider a system with $ N $ qubits.
We divide the system into two regions (i.e., two groups of qubits), written as $ X $ and $ X^{\mathrm{c}} $, each containing $ n $ and $ N-n $ qubits.
We denote the dimension of the Hilbert space of $ X $ as $ d_X:=2^{n} $.
The partial cyclic permutation operator $ C_X $ associated with $X$ is defined as the operator that cyclically permutes the $ k $ replicas of  subsystem $ X $ as
\begin{align}
C_X\ket{x_1y_1, x_2y_2,\dots, x_ky_k}
=
\ket{x_2y_1, x_3y_2,\dots, x_1y_k},
\label{eq:partial_cycle}
\end{align}
where $ \ket{x_i} $ and $ \ket{y_i} $ ($ x_i\in \{0,1\}^n,\  y_i\in \{0,1\}^{N-n}$, and $ i=1,\dots, k $) respectively represent the computational bases of $X$ and $X^{\mathrm{c}}$.
We note that $C_X $ cannot be represented as the tensor product form $ A^{\otimes k} $, and its support is in the entire Hilbert space but not in the energy shell $ \Hsh^{\otimes k} $.
With the use of $ C_X $, the $ k $-REE of $ \ket{E_i} $ is now defined as
\begin{align}
R_i^{(k)}(X):=\frac{1}{1-k}\log \braket{E_i\dots E_i|C_X|E_i\dots E_i},
\label{eq:def_k-REE}
\end{align}
which is equivalent to the standard definition through the reduced density operator~\cite{Eisert2010}.
An important point here is that the $ k $-REE is expressed as the expectation value of $ C_X $, which allows us to apply the $ k $-ETH~\eqref{eq:k-ETH}.

\subsection{Page curve
\label{sec:Entanglement:Infinite}}

We first consider the Page curve of energy eigenstates at infinite temperature, which has been numerically studied in Refs.~\cite{Garrison2018,Beugeling2015JSM,Vidmar2017}.
In this subsection, we simplify our setting by considering the entire Hilbert space of $ N $ qubits  (including negative temperatures) with dimension $d:= 2^N$, instead of the energy shell~\eqref{eq:shell}.
For further simplicity, we consider the 2-REE, where the partial cyclic permutation operator is reduced to the partial swap operator $ S_X $, which acts as
\begin{align}
S_X\ket{x_1y_1, x_2y_2}
=
\ket{x_2y_1, x_1y_2}
\label{eq:partial_swap}.
\end{align}
In this case, the 2-ETH for $ S_X $ is expressed as
\begin{align}
\braket{E_iE_i|S_X|E_iE_i}
&=\frac{d/d_X+d_X}{d+1}
\label{eq:2-ETH_1_swap},
\\
\braket{E_iE_j|S_X|E_iE_j}
&=\frac{d^2/d_X-d_X}{d^2-1}
\label{eq:2-ETH_2_swap},
\\
\braket{E_iE_j|S_X|E_jE_i}
&=\frac{dd_X-d/d_X}{d^2-1},
\label{eq:2-ETH_3_swap}
\end{align}
where we used
$ \tr{S_X}=d^2/d_X $
and
$ \tr{SS_X}=dd_X $.

If $ X $ is much smaller than the entire system
($ n\ll N $),
the second terms of the numerators on the right-hand sides of
Eqs.~\eqref{eq:2-ETH_1_swap} and \eqref{eq:2-ETH_2_swap}
are negligible,
and the 2-ETH is approximated as
\begin{align}
\braket{E_iE_i|S_X|E_iE_i}
&\approx d_X^{-1},
\\
\braket{E_iE_j|S_X|E_iE_j}
&\approx d_X^{-1},
\\
\braket{E_iE_j|S_X|E_jE_i}
&\approx 0.
\end{align}
The diagonal elements of $ S_X $ are nonzero and take the same value for all $ i, j$,
whereas the off-diagonal elements are approximately zero,
which is the same as the case of  the 2-ETH for $ A^{\otimes 2} $ (see Eqs.~\eqref{eq:2-ETH_1_product_op}-\eqref{eq:2-ETH_3_product_op}).

On the other hand,
when $ X $ is a half of the system ($ n=N/2 $),
the second term of the numerator on the right-hand side of
Eq.~\eqref{eq:2-ETH_1_swap}
are comparable to the first term.
In this case,
the 2-ETH is approximated as
\begin{align}
\braket{E_iE_i|S_X|E_iE_i}
&\approx 2d^{-1/2}
\label{eq:2-ETH_1_swap_half},
\\
\braket{E_iE_j|S_X|E_iE_j}
&\approx  d^{-1/2}
\label{eq:2-ETH_2_swap_half},
\\
\braket{E_iE_j|S_X|E_jE_i}
&\approx  d^{-1/2}
\label{eq:2-ETH_3_swap_half}.
\end{align}
The factor of 2 in front of $ d^{-1/2} $ in Eq.~\eqref{eq:2-ETH_1_swap_half} is important, which leads to the Page correction as shown below.

From Eqs.~\eqref{eq:def_k-REE} and~\eqref{eq:2-ETH_1_swap},
the 2-REE of $ \ket{E_i} $ is now obtained as a consequence of the 2-ETH:
\begin{align}
R^{(2)}_{i}(X)
&=
-\log \frac{d/d_X+d_X}{d+1}
\\
&=
n\log 2
-\log\left(
1+
\frac{2^{2n}-1}{2^{N}+1}
\right).
\label{eq:Page_curve}
\end{align}
The first term in the second line represents the volume law, and
the second term is the subleading correction to the volume law.
The correction term takes its maximum value $  \log 2 $ at $ n=N/2 $,
while negligible at $ n\ll N $.
Equation~\eqref{eq:Page_curve} as a function of $ n $
is called the (2-\renyi) Page curve~\cite{Page1993}.

The volume law of the 2-REE can also be accounted for by the 1-ETH,
while the Page correction cannot~\cite{Garrison2018}.
To see this,
we set $ k=2 $ and  $ O=S_X $ in Eq.~\eqref{eq:non-interacting_k-ETH},
and obtain
\begin{align}
\braket{E_iE_i|S_X|E_iE_i}=d_X^{-1}
\label{eq:non-interacting_2-ETH},
\end{align}
which is a straightforward generalization of the 1-ETH to the partial swap operator.
It is obvious that Eq.~\eqref{eq:non-interacting_2-ETH}
gives the volume law,
but not the Page correction.
Thus,
the Page correction for individual energy eigenstates
comes essentially from the 2-ETH.

We note that Eqs.~\eqref{eq:2-ETH_2_swap} and \eqref{eq:2-ETH_3_swap} are not directly related to the entanglement entropy of energy eigenstates.
These equations represent the correlations among eigenstates and are closely related to the 2-OSEE of the LTE.
Especially, Eqs.~\eqref{eq:2-ETH_2_swap} and \eqref{eq:2-ETH_3_swap} give the volume law and the Page correction of the 2-OSEE, respectively (see Appendix~\ref{sec:OSEE}).

\subsection{Thermal Page curve
\label{sec:Entanglement:Thermal}}
We next consider the 2-REE for the finite temperature case, where the Hilbert space $\Hsh$ does not equal the entire Hilbert space, but is spanned by energy eigenstates only in the energy shell~\eqref{eq:shell}.
In this case, the support of $ S_X $ is not in the energy shell $ \Hsh^{\otimes 2} $,
and thus we need to project $ S_X $ to $  \Hsh^{\otimes 2} $.
We denote the projection operator from the entire Hilbert space to $ \Hsh $ by $ \Psh $
and the projected partial swap operator
by $ \tilde{S}_X:= \Psh^{\otimes 2}S_X\Psh^{\otimes 2}$.
Then, we can easily show that
\begin{align}
	\tr{\tilde{S}_X}
	&=\Dsh^2 e^{-R_{\mathrm{mc}}^{(2)}(X)},
	\label{eq:ppswap_trace_1}
	\\
	\tr{S\tilde{S}_X}
	&=\Dsh^2 e^{-R_{\mathrm{mc}}^{(2)}(X^{\mathrm{c}})}.
	\label{eq:ppswap_trace_2}
\end{align}
Here,
$ R_{\mathrm{mc}}^{(2)}(X) $ is 
the 2-\renyi entropy of the reduced microcanonical state on $ X $,
defined by $ R_{\mathrm{mc}}^{(2)}(X):=
-\log \tr{\rho_{\mathrm{mc}}^{\otimes 2}S_X} $.
From Eqs.~\eqref{eq:ppswap_trace_1} and~\eqref{eq:ppswap_trace_2},
we find that the 2-ETH for $ \tilde{S}_X $ is expressed as
\begin{align}
\braket{E_iE_i|\tilde{S}_X|E_iE_i}
&=\frac{\Dsh}{\Dsh+1}\left(
e^{-R_{\mathrm{mc}}^{(2)}(X)}+e^{-R_{\mathrm{mc}}^{(2)}(X^\mathrm{c})}
\right),
\label{eq:2-ETH_1_swap_thermal}
\\
\braket{E_iE_j|\tilde{S}_X|E_iE_j}
&=\frac{\Dsh^2}{\Dsh^2-1}\left(
e^{-R_{\mathrm{mc}}^{(2)}(X)}-e^{-R_{\mathrm{mc}}^{(2)}(X^\mathrm{c})}/\Dsh
\right),
\label{eq:2-ETH_2_swap_thermal}
\\
\braket{E_iE_j|\tilde{S}_X|E_jE_i}
&=\frac{\Dsh^2}{\Dsh^2-1}\left(
e^{-R_{\mathrm{mc}}^{(2)}(X^\mathrm{c})}-e^{-R_{\mathrm{mc}}^{(2)}(X)}/\Dsh
\right).
\label{eq:2-ETH_3_swap_thermal}
\end{align}
These are the extensions of Eqs.~\eqref{eq:2-ETH_1_swap}-\eqref{eq:2-ETH_3_swap} to finite temperature.

Taking the logarithm of Eq.~\eqref{eq:2-ETH_1_swap_thermal},
we obtain the following relation for the 2-REE:
\begin{align}
R^{(2)}_{i}(X)&=R_{\mathrm{mc}}^{(2)}(X)
-\log\frac{ 1+e^{R_{\mathrm{mc}}^{(2)}(X)-R_{\mathrm{mc}}^{(2)}(X^\mathrm{c})}}{1+1/\Dsh}
\label{eq:thermal_Page_curve}
\\
&=:R_{\mathrm{th}}^{(2)}(X),
\end{align}
which is  the finite-temperature Page curve.
The first term on the right-hand side of Eq.~\eqref{eq:thermal_Page_curve}
represents the local thermality, which is accounted for by the 1-ETH~\eqref{eq:non-interacting_k-ETH}.
The second term reflects the correlations among energy eigenstates, and represents the finite-temperature counterpart of the Page correction, and becomes $ \log 2 $ at $ n=N/2 $, which is the same value as the Page correction obtained in Sec.~\ref{sec:Entanglement:Infinite}.
We note that essentially the same formulas as Eq.~\eqref{eq:thermal_Page_curve} have been obtained in Ref.~\cite{Nakagawa2018,Fujita2018,Lu2019} by using random pure states.

We next show that $ R_{\mathrm{th}}^{(2)}(X) $ follows a universal formula of the 2-REE of thermal pure states, which has been proposed in Refs~\cite{Nakagawa2018,Fujita2018}.
It is written as
\begin{align}
R_{\mathrm{th}}^{(2)}(X)=n\log a(\beta)-\log (1+a(\beta)^{-N+2n})+\log K(\beta),
\label{eq:universal_form}
\end{align}
where $ a(\beta) $ and $ K(\beta) $ are constants depending only on the inverse temperature $ \beta $.
When the interaction of the system is spatially homogeneous, we expect that the reduced microcanonical state of $ X $ gives the volume law of the 2-REE.
We thus assume that there exists $ 1\leq w(u)\leq 2 $ depending only on the energy density $ u:=E/N $, and that the following relation holds:
\begin{align}
R_{\mathrm{mc}}^{(2)}(X)
= n\log w(u).
\label{eq:volume-law}
\end{align}
By substituting Eq.~\eqref{eq:volume-law} into Eq.~\eqref{eq:thermal_Page_curve},
we obtain
\begin{align}
R_{\mathrm{th}}^{(2)}(X) =
n\log w(u) - \log (1+w(u)^{-N+2n})
+\mathcal{O}(\Dsh^{-1}).
\label{eq:thermal_Page_curve_2}
\end{align}
It is known that
there is a one-to-one correspondence between
the energy density and the inverse temperature except for  phase transition points~\cite{Ruelle1999}.
Assuming no phase transition,
we can rewrite $w$ as a function of $\beta$.
Then,
we rewrite Eq.~\eqref{eq:thermal_Page_curve_2} as
\begin{align}
R_{\mathrm{th}}^{(2)}(X)& =
n\log w(\beta) - \log (1+w(\beta)^{-N+2n})
+\mathcal{O}(\Dsh^{-1}).
\label{eq:thermal_Page_curve_3}
\end{align}
This expression coincides with the universal form~\eqref{eq:universal_form}
by setting $ a(\beta)=w(\beta) $ and $ \log K(\beta) =\mathcal{O}(\Dsh^{-1})$.
As mentioned before,
the original formula~\eqref{eq:universal_form} has been obtained
by calculating the 2-REE of random pure states.
On the other hand,
we derived Eq.~\eqref{eq:thermal_Page_curve_3} for energy eigenstates  from the 2-ETH for the partial swap operator. 

We consider the infinite temperature case, $ \beta \to 0 $, where the energy shell is given by Eq.~\eqref{eq:shell}.
In this case, we can expect that for $ 1\leq n\leq N/2 $
\begin{align}
R_{\mathrm{mc}}^{(2)}(X)=n\log 2+o(1).
\label{eq:maximum_EE}
\end{align}
This is rigorously provable for $n \ll N$ from the equivalence of ensembles~\cite{Ruelle1999,Tasaki2018}.
Under the assumption~\eqref{eq:maximum_EE}, Eq.~\eqref{eq:thermal_Page_curve} reduces to the Page curve~\eqref{eq:Page_curve} of the entire Hilbert space of dimension $2^N$, up to the $o(1)$ correction (see Eq.~\eqref{eq:Page_curve} in Sec.~\ref{sec:Entanglement:Infinite})..
We note that the $ o(1) $ correction term in Eq.~\eqref{eq:maximum_EE} completely vanishes if we take the entire Hilbert space.

\subsection{$ k $-REE}

The foregoing argument can apply to the higher $ k $-REE ($ k\geq 3 $).
We consider the $ k $-ETH~\eqref{eq:k-ETH} for the projected partial cyclic permutation operator $ \tilde{C}_X:= \Psh^{\otimes k}C_X\Psh^{\otimes k} $.
In this case, the $ k $-ETH with $ i_1=i_2=\cdots =i_k=:i $ is rewritten as
\begin{align}
&\braket{E_{i}\dots E_{i}| \tilde{C}_X |E_{i}\dots E_{i}}
\nonumber
\\
&=
\frac{\Dsh^{k-1}}{(\Dsh+1)\cdots (\Dsh+k-1)} \sum_{\tau \in S_k}
\tr{W_{\tau}C_X \rho_{\mathrm{mc}}^{\otimes k}},
\label{eq:k-ETH_cycle}
\end{align}
where we used $ \sum_{\pi\in S_k}\Wg_{\pi,\tau}(\Dsh)=\Dsh^{-1}(\Dsh+1)^{-1}\cdots (\Dsh+k-1)^{-1} $~\cite{Gu2013}.
The leading term of the right-hand side of Eq.~\eqref{eq:k-ETH_cycle} comes from the term of $ \tau =I $, and is given by
\begin{align}
\tr{C_X \rho_{\mathrm{mc}}^{\otimes k}}
=
e^{-(k-1) R_{\mathrm{mc}}^{(k)}(X)},
\end{align}
where $ R_{\mathrm{mc}}^{(k)}(X):=(1-k)^{-1}\log \tr{C_X \rho_{\mathrm{mc}}^{\otimes k}} $ is the $ k $-\renyi entropy of the reduced microcanonical state on $ X $.
For $ \Dsh\gg 1 $, Eq.~\eqref{eq:k-ETH_cycle} is approximated as
\begin{align}
\braket{E_{i}\dots E_{i}| \tilde{C}_X |E_{i}\dots E_{i}}
\approx
e^{-(k-1) R_{\mathrm{mc}}^{(k)}(X)}+ w_k(X),
\end{align}
where $w_k(X):= \sum_{\tau \neq I}\tr{W_{\tau}C_X \rho_{\mathrm{mc}}^{\otimes k}} $ is the subleading term.
Taking the logarithm, we obtain the following relation for the $ k $-REE:
\begin{align}
R_i^{(k)}(X)
&\approx  R_{\mathrm{mc}}^{(k)}(X)
+
\frac{1}{1-k} \log\left(1+
e^{(k-1) R_{\mathrm{mc}}^{(k)}(X)}w_k(X)
 \right)
\label{eq:thermal_k-REE}\\
&=: R_{\mathrm{th}}^{(k)}(X).
\end{align}
As in Eq.~\eqref{eq:thermal_Page_curve}, the first term on the right-hand side of Eq.~\eqref{eq:thermal_k-REE} represents the local thermality, and the second term is regarded as the generalized Page correction.
Thus, $ R_{\mathrm{th}}^{(k)}(X) $ is a higher-order counterpart of  the finite-temperature Page curve~\eqref{eq:thermal_Page_curve}.
For example, in the case of $ k=3 $, we have
\begin{align}
R_{\mathrm{th}}^{(3)}(X)=
 R_{\mathrm{mc}}^{(3)}(X)
-
\frac{1}{2} \log\left(1+
e^{2R_{\mathrm{mc}}^{(3)}(X)}w_3(X)
\right),
\label{eq:thermal_3-REE}
\end{align}
where
\begin{align}
w_3(X)
&=e^{-2 R_{\mathrm{mc}}^{(3)}(X^{\mathrm{c}})}
+
3\tr{\rho_{\mathrm{mc}}(\rho^X_{\mathrm{mc}}\otimes \rho^{X^\mathrm{c}}_{\mathrm{mc}})}
\nonumber\\
&\span+
\tr{C_XC_{X^{\mathrm{c}}}^{-1}\rho_{\mathrm{mc}}^{\otimes 3}},
\end{align}
and $ \rho^X_{\mathrm{mc}} $ is the reduced microcanonical state on $ X $.
Again, essentially the same formulas as Eq.~\eqref{eq:thermal_3-REE} have been obtained in Ref.~\cite{Nakagawa2018,Fujita2018,Lu2019} by using random pure states.
In contrast to the case of $ k=2 $,  the finite-temperature Page correction of the 3-REE  (i.e., the second term of Eq.~\eqref{eq:thermal_3-REE}) does not match the original value of the infinite-temperature Page correction even at $ n=N/2 $.

We remark on the $ 1 $-REE, namely the von Neumann entanglement entropy
$ R^{(1)}(X) :=-\tr{\rho_X\log\rho_X}$.
In contrast to the case of $k \geq 2$, the Page curve of the 1-REE does not follow from the 1-ETH, because $ R^{(1)}(X)  $ cannot be expressed as the expectation value of an operator that is independent of the state.
However, the 2-ETH gives the following lower bound on the $ 1 $-REE of the energy eigenstate:
\begin{align}
R_i^{(1)}(X)\geq R_i^{(2)}(X)=R_{\mathrm{th}}^{(2)}(X),
\end{align}
which ensures the volume law of the 1-REE.

Recent studies~\cite{LeBlond2019,Vidmar2017quad,Vidmar2018,Hackl2019} showed that the average of the 1-REE over all the energy eigenstates of an integrable system obeys the volume law,
but does not follow the Page correction in a non-local region.
Their results suggest that the average error of the 2-ETH for the non-local partial swap operator quantified by Eq.~\eqref{eq:2-ETH_1_error} does not decay faster than $ d^{-1/2} $.

\section{Numerical verifications of the 2-ETH
	\label{sec:Numeric}}

By using numerical exact diagonalization, we confirm that the approximate 2-ETH is true for the one-dimensional XXZ ladder model, which is nonintegrable except for a particular parameter point~\cite{Beugeling2014,Beugeling2015,Beugeling2015JSM}.
We show that the indicator of the 2-ETH~\eqref{eq:2-ETH_indicator} decays polynomially as $\Dsh^{-a}$ with $a \sim 0.5$ for operators in the form of $ O=A^{\otimes 2} $.
We also show that the indicator decays as $d^{-a}$ with $a \sim 0.5$ for the local partial swap operators, while decays with $a \sim 1$ for the non-local (half) partial swap operator.
The latter is in particular consistent with the random operator argument~\eqref{eq:typical_error}, and makes the Page correction non-negligible.  

\subsection{Model and Method
\label{sec:Numeric:models}}

\begin{figure}
	\centering
	\includegraphics[width=0.6\linewidth]{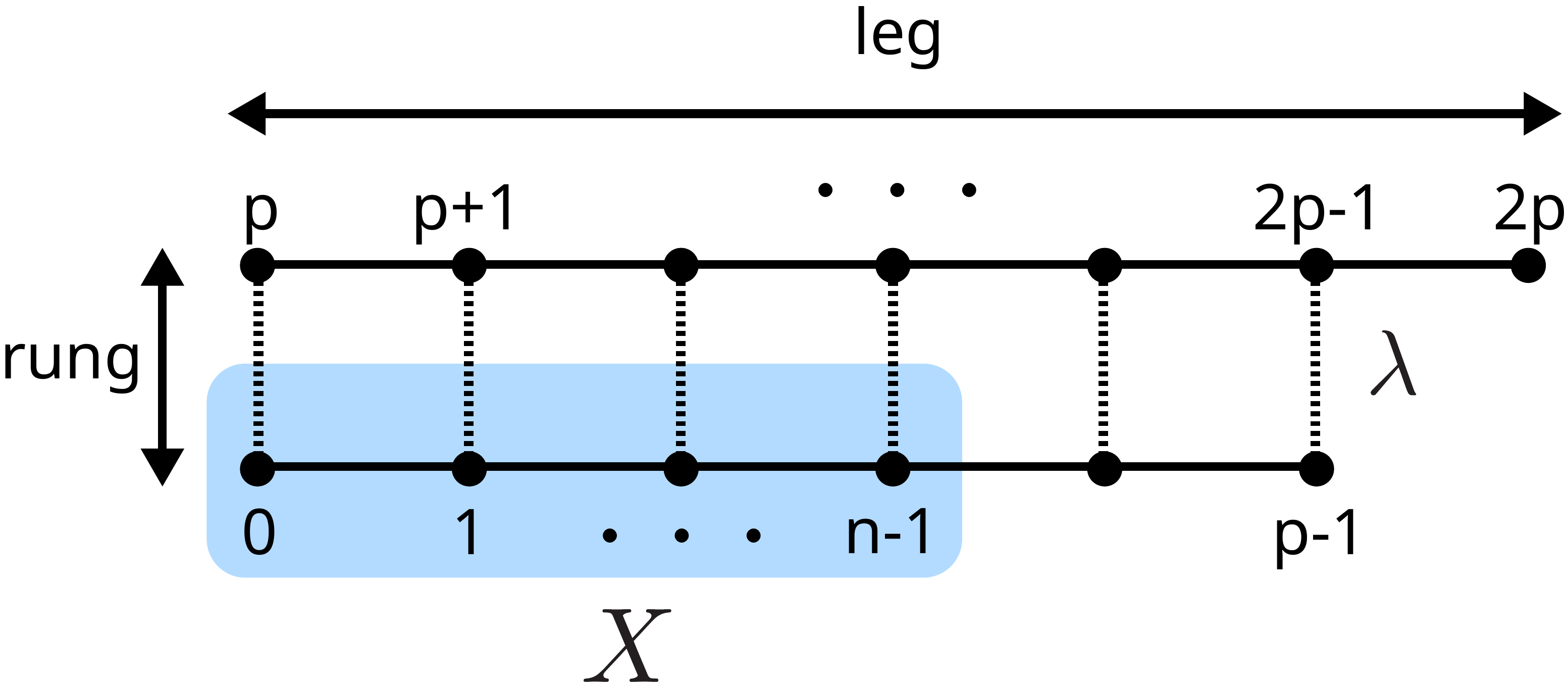}
	\caption{\label{fig:ladder}
		The graph structure of the XXZ ladder model~\eqref{eq:XXZ-ladder} and the labeling of sites.
		The blue region shows  $ X:=\{0,1,\dots, n-1\} $, which is used in the calculation of $ I_2(S_X) $.}
\end{figure}

We consider the one-dimensional spin-1/2 Heisenberg XXZ ladder model, which is composed by two chains whose lengths are $ p $ and $ p+1 $ respectively
 (Fig.~\ref{fig:ladder}).  
The total number of spin is $ N=2p+1 $.
The Hamiltonian is described by
\begin{align}
 H_\mathrm{XXZ}
 &:=
 \sum_{i=0}^{p-2} h_{i,i+1}
 +
 \sum_{i=p}^{2p-1} h_{i,i+1}
 +
 \lambda\sum_{i=0}^{p-1}  h_{i,i+p},
 \label{eq:XXZ-ladder}
 \\
 h_{i,j}&:=X_iX_j+Y_i Y_j+\Delta Z_iZ_j,
\end{align}
where $ N $ is the number of sites and $X_i, Y_i, Z_i$ are the Pauli operators at site $ i $.
As the number of up-spins 
$ N_\uparrow:=\sum_{i=1}^N (Z_i+1)/2  $ conserves
$[H_\mathrm{XXZ},N_\uparrow]=0$,
we use the sector of $ N_\uparrow=p $ in the following calculations.
The anisotropy parameter $ \Delta  $ is fixed as $  \Delta=0.8 $.
The coupling constant $ \lambda $  in the rungs tunes the integrability of the model.
The model is nonintegrable at $ \lambda\neq 0 $ and is integrable at $ \lambda=0 $.

We numerically calculate the indicator of the 2-ETH $  I_2(O) $ defined in Eq.~\eqref{eq:2-ETH_indicator} by using numerical exact diagonalization.
We sample all of the energy eigenstates whose energies are in the energy shell $[E-\Delta E, E] $, whose Hilbert-space dimension is written as  $ \Dsh $.
To remove the finite-size effect and the statistical error, we paid attention on the following points in line with Refs.~\cite{Beugeling2014,Yoshizawa2018}.
It is desirable that the width $ \Delta E $ is large, because the number of eigenstates in $[E-\Delta E, E]$ is responsible for the statistical error of sampling. 
We thus adopt $ \Delta E=N \delta$ with $\delta$ being a fixed constant. 
On the other hand, the matrix elements of an operator $ \braket{E_iE_j|O|E_kE_l} $ sensitively depend on the eigenenergies.
Thus, for $\Delta_i^1 (O)$, we calculate the microcanonical ensemble average
by using a thinner  energy shell $ [E_i-\delta_{\mathrm{mc}} , E_i] $
by changing $E_i$ for each energy eigenstate $\ket{E_i} $,
where $\delta_{\rm mc}$ is an $N$-independent constant that is much smaller than $\Delta E$.
We also calculate the microcanonical ensemble average for $  \Delta_{ij}^{2}(O), \Delta_{ij}^{3}(O) $ by using the energy shell
$ [E_{ij}-\delta_{\mathrm{mc}}, E_{ij}] $ with $ E_{ij}:=(E_i+E_j)/2$.
We note that  we need to project $ O $ to the Hilbert space $ \Hsh $ of the energy shell to calculate the indicator of the 2-ETH $  I_2(O) $.

In the following calculations,
we set  $ E $ to the energy at infinite temperature $ E=\tr{H_{\mathrm{XXZ}}}/2^N $, and set $ \delta=0.02$ and $ \delta_{\mathrm{mc}}=0.1 $.

\subsection{Tensor product operator
	\label{sec:Numeric:product}}

\begin{figure}
	\includegraphics[width=1.0\linewidth]{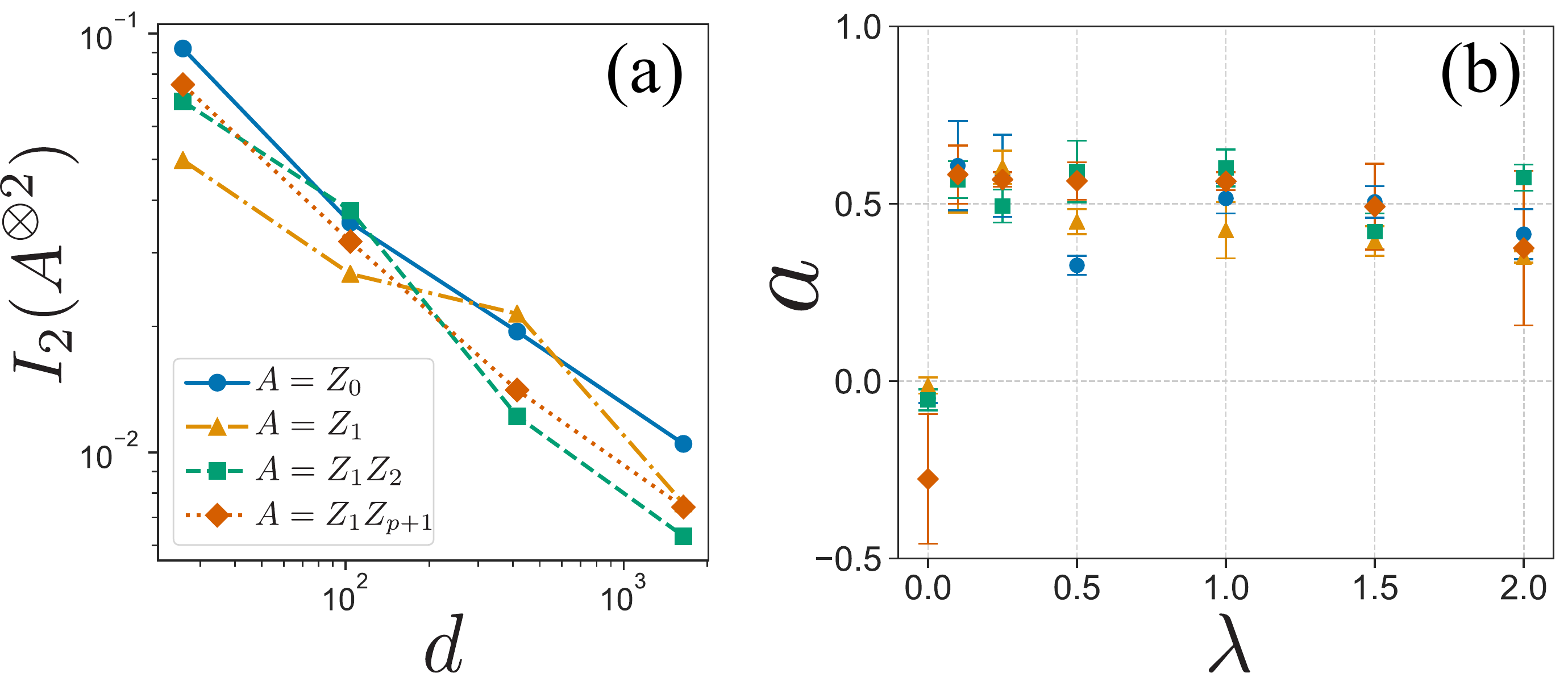}
	\caption{\label{fig:tensor}
		(a) The $\Dsh$-dependence of $I_2(A^{\otimes 2})$ with $ \lambda =1 $ (nonintegrable).
		$ I_2(A^{\otimes 2} ) $ decays polynomially with respect to $ \Dsh $.
		(b) The $\lambda$-dependence of the exponent $ a $ of $  I_2(A^{\otimes 2})=\mathcal{O}(\Dsh^{-a})  $.
		At $ \lambda\neq 0 $ (nonintegrable), the exponent $ a $ is positive and close to $ 0.5 $.
		At $ \lambda=0 $ (integrable), $a$ is around zero and seems even negative but almost within the numerical error.
		In both panels, the circles, triangles, squares, and diamonds represent the results
		for  $ Z_0 $, $ Z_1$, $ Z_1Z_2$, and $Z_1Z_{p+1} $, respectively.
	}
\end{figure}

We show the $ \Dsh $-dependence of the 2-ETH indicator~\eqref{eq:2-ETH_indicator} for the tensor product operator $O=A^{\otimes 2}$ with $ A\in\bop(\Hsh) $.
In the following, we take $Z_0, Z_1, Z_1Z_2,$ and $ Z_1Z_{p+1}$ for $A$. 
We remark that the traces of the above operators on the entire Hilbert space are exactly zero,
while their microcanonical averages are not necessarily zero even at infinite temperature due to the present definition of the energy shell~\eqref{eq:shell}, as mentioned in Sec.~\ref{sec:k-ETH:Approximate_k-ETH}.

Figure~\ref{fig:tensor}~(a) shows the $\Dsh$-dependence of $I_2(A^{\otimes 2})$
for the XXZ ladder model \eqref{eq:XXZ-ladder} with $\lambda=1$. 
The system size is $ N=11,13,15,17 $. 
For all of the four cases, $I_2(A^{\otimes 2})$ decays polynomially with respect to $ \Dsh $. 
This result implies that the approximate 2-ETH holds in this model.

To investigate the $d$-dependence of $ I_2(A^{\otimes 2}) $ in more detail,
we fit the numerical data of $\log  I_2(A^{\otimes 2} ) $ against a fitting function 
$f(\Dsh):=-a\log \Dsh +b$ with fitting parameters $a$ and $b$. 
The positive $a$ implies the approximate  2-ETH.
We perform the fitting analysis by changing the rung interaction $\lambda$
to study the role of the nonintegrability.
Figure~\ref{fig:tensor}~(b) shows the $\lambda$-dependence of
the exponent  $a$.
The error bars show the numerical error of the fitting.
While $ a\sim 0 $  in the integrable case ($\lambda=0$), 
$a$ takes a finite positive value close to $0.5$ in the nonintegrable cases ($\lambda\neq 0$). 
This implies that $  I_2(A^{\otimes 2}  )$ does not decay for the integrable case, while decays for the nonintegrable cases.

If the Hamiltonian is sufficiently quantum chaotic, $  I_2(A^{\otimes 2}  )$ would match the prediction~\eqref{eq:scaling_RMT_product} of random matrix theory.
Since we numerically find that the microcanonical average of $ A $ scales as $ \braket{A}_{\mathrm{mc}}= \mathcal{O}(\Dsh^{-\alpha}) $ with $ \alpha\sim 0.1 $ (see Appendix.~\ref{sec:mc_exponent}), Eq.~\eqref{eq:scaling_RMT_product}  gives $  I_2(A^{\otimes 2}) = \mathcal{O}(\Dsh^{-a})$ with $ a\sim 0.6 $.
On the other hand, Fig.~\ref{fig:tensor}~(b) shows that the exponent takes $ a\sim 0.5 $ in the nonintegrable cases,
which is a bit smaller than the prediction.
This would reflect the finite-size effect and perhaps the fact that the Hamiltonian is not perfectly chaotic due to the locality of interactions.

\subsection{Partial swap operator
\label{sec:Numeric:swap}}

\begin{figure}
	\includegraphics[width=1.0\linewidth]{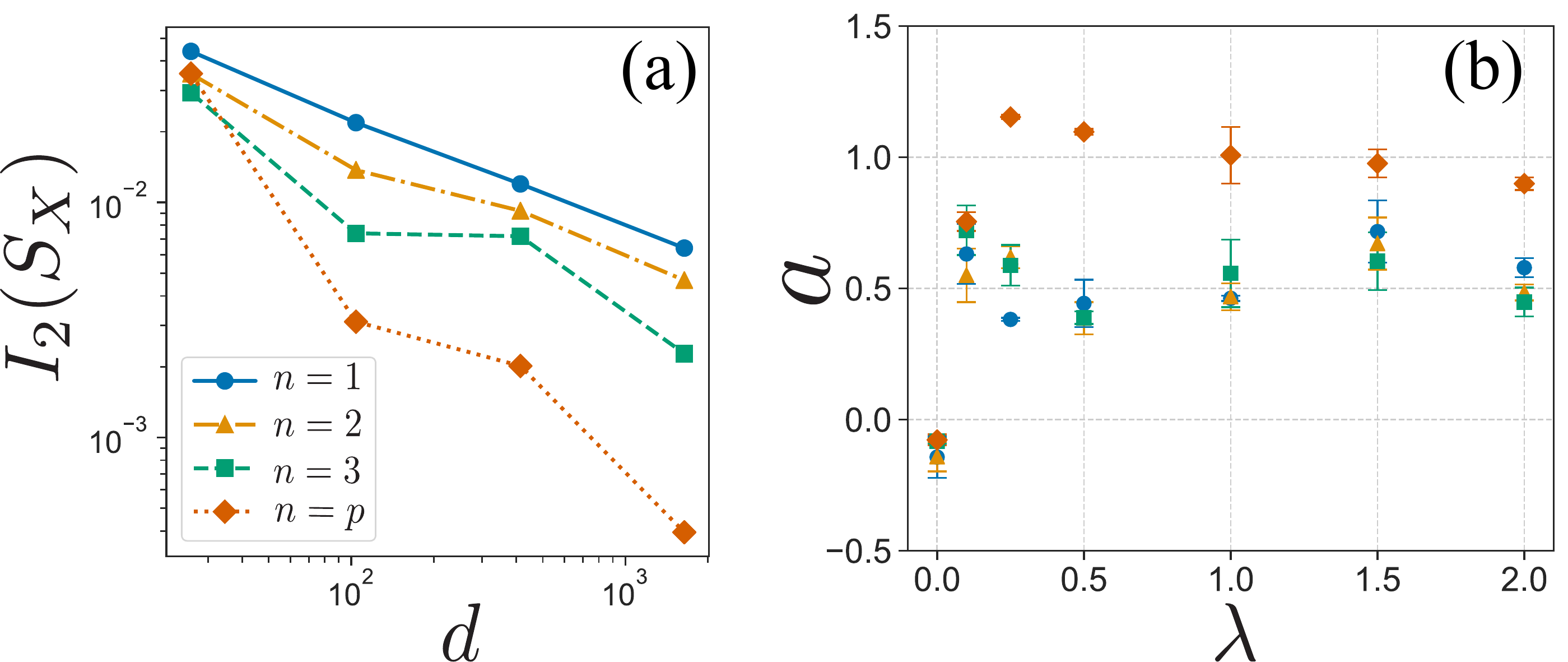}
	\caption{\label{fig:swap}
		(a) The $\Dsh$-dependence of $I_2(S_X)$ with $ \lambda=1 $.
		$ I_2(S_X ) $ decays polynomially with respect to $ \Dsh $.
		(b) The $\lambda$-dependence of the exponent $ a $ of $  I_2(S_X) =\mathcal{O}(\Dsh^{-a}) $.
		At $ \lambda\neq 0 $ (nonintegrable), $ a $ is positive, especially  $a\sim 0.5 $ for $ n=1, 2, 3 $, and $ a\sim1 $ for $ n=p $.
		As is the case for Fig.~\ref{fig:tensor}~(b),
		$ I_2(S_X)  $ does not decay at $ \lambda=0 $ (integrable).
		In both panels, the circles, triangles, squares and diamonds represent the results
		for  $ n=1,2,3$, and $ n=p $.
	}
\end{figure}

We next show the $ \Dsh $-dependence of $I_2$ for the partial swap operator $S_X$
defined in Eq.~\eqref{eq:partial_swap}.
The region $X$ is chosen as $X=\{0,1,\dots, n-1\}$ (see Fig.~\ref{fig:ladder}).
Figure~\ref{fig:swap}~(a) shows the $ \Dsh $-dependence of $ I_2(S_X) $.
The parameters are the same as those in Fig.~\ref{fig:tensor}.
Similar to the case of $ O=A^{\otimes2} $,
$I_2(S_X)$ decays polynomially with respect to $ \Dsh $, and the approximate 2-ETH holds. 
Thus, all the eigenstates obey the volume law of the 2-REE.

We also perform the fitting analysis in the same way as in Sec.~\ref{sec:Numeric:product}. 
Figure~\ref{fig:swap}~(b) shows the $\lambda$-dependence of $a$.
In the integrable case ($\lambda=0$),
the exponent $a$ takes a value around zero but slightly negative. 
The negativity could be a finite-size effect.
Thus, the approximate 2-ETH does not hold in the integrable case. 
On the other hand, in the nonintegrable cases ($\lambda\neq 0$), the exponent $ a $ takes a sufficiently large positive value, which implies that the approximate 2-ETH holds.

Let us consider the implication of the above result in more detail.
For $n=1,2,3$, we find that  $a\sim 0.5$ in the nonintegrable cases ($\lambda\neq 0$), which is close to the case of $ O=A^{\otimes 2} $.
This is understood from the fact that $ S_X $ is represented as a sum of operators in the tensor product form:
\begin{align}
S_X= \frac{1}{d_X}\sum_i P_i\otimes P_i.
\end{align}
Here, the summation is taken over all many-body Pauli operator $ P_i $ supported on $ X $,  whose number $d_X^2$ is independent of $d$ as long as $X$ is local.
Thus, we can apply Eq.~\eqref{eq:scaling_RMT_product} to this case.
On the other hand, for $n=p=(N-1)/2$, the exponent of the half-swap operator $S_X$ takes $ a\sim 1 $.
In this case, $d_X= \mathcal{O}(\Dsh^{1/2})$ grows with $d$, and thus Eq.~\eqref{eq:scaling_RMT_product}  cannot apply.
Instead, Eq.~\eqref{eq:typical_error} is now consistent with the numerically obtained $a$, suggesting that the half-swap operator is a typical operator in $\bop(H^{\otimes 2})$.
Thanks to this scaling of the error of the 2-ETH, the Page correction discussed in Sec.~\ref{sec:Entanglement} becomes non-negligible in the 2-REE.

We note that in our calculations of the second terms
on the right-hand sides of Eqs.~\eqref{eq:2-ETH_1_error}-\eqref{eq:2-ETH_3_error},
we used random states sampled from the Hilbert space corresponding to the energy shell $ [E_i-\delta_{\mathrm{mc}} , E_i] $ for Eq.~\eqref{eq:2-ETH_1_error} and  $[E_{ij}-\delta_{\mathrm{mc}}, E_{ij}] $ for Eqs.~\eqref{eq:2-ETH_2_error}, \eqref{eq:2-ETH_3_error}, instead of taking the trace directly.
It is known that the expectation value over such a random state agrees with the microcanonical ensemble with the probability close to unity, if the number of energy eigenstates in the energy shell is sufficiently large~\cite{Popescu2006}.

\subsection{Summary of numerics
	\label{sec:Numeric:summary}}

\begin{table}
	\centering
	\caption{\label{table:summary}
		Summary of the exponent $ a $ of the 2-ETH indicator $I_2 (O) = \mathcal{O}(\Dsh^{-a})$.
		The approximate 2-ETH holds only in the nonintegrable cases.
	}
	\setlength{\tabcolsep}{3mm} 
	\renewcommand{\arraystretch}{1.5}
	\begin{tabular}{l|c|c|c}
		\hline\hline
		& $A^{\otimes 2}$  & \multicolumn{2}{c}{$S_X$}        \\  \cline{3-4}
		& & local             & non-local                   \\ \hline
		Integrable                      & $ \sim0$           & $  \sim0$         & $ \sim0$                   \\\hline
		Nonintegrable & $\sim 0.5$ & $ \sim 0.5$ & $ \sim1 $ \\
		\hline\hline
	\end{tabular}
\end{table}

Table~\ref{table:summary} summarizes our numerical results about the exponent $ a $ of $I_2(O)= \mathcal{O}(\Dsh^{-a})$.
In the integrable case, $I_2(O)$ does not decay and thus the approximate 2-ETH fails.
On the other hand,  $I_2(O)$ decays polynomially with respect to $\Dsh$ in the nonintegrable cases, implying  that the approximate 2-ETH is true. 
For the local operators such as $A^{\otimes2}\ (A=Z_0, Z_1, Z_1Z_2, Z_1Z_{p+1})$ and $S_X$ $(n=1,2,3)$, the exponent satisfies $a\sim 0.5$,
which is close to (but does not perfectly match) the prediction of random matrix theory discussed in Sec.~\ref{sec:k-ETH:Approximate_k-ETH}.
For the non-local swap operator $S_X\ (n=p)$,  the exponent satisfies $a\sim 1$,
which is consistent with the typicality prediction~\eqref{eq:typical_error}.

As is the case for the conventional ETH, we can also consider the weaker version of the 2-ETH, which states that the fraction of the eigenstates that do not satisfy the 2-ETH vanishes in the thermodynamic limit. 
We expect that weak 2-ETH is true even in the integrable case, as is the case of the weak 1-ETH~\cite{Biroli2010,Mori2016,Iyoda2017}.
However, this is not clear from our numerical data because of the finite-size effect.

\section{Partial unitary design
	\label{sec:PU-design}}
In this section, we reconsider the $ k $-ETH in light of quantum pseudo-randomness in order to give an information-theoretic view on the hierarchy of quantum many-body chaos.
We introduce a slight generalization of unitary $ k $-designs, named partial unitary (PU) $ k $-designs.
A  PU $ k $-design is defined as an ensemble of unitaries that is indistinguishable from the HRU as long as one only observes a limited class of observables.
The formula of the $ k $-ETH~\eqref{eq:k-ETH} is obtained from a PU $ k $-design for the LTE.

\subsection{Basic idea}

The essence of the $ k $-ETH is quantum pseudo-randomness of chaotic dynamics.
In the quantum information context, quantum pseudo-randomness is formalized by the concept of unitary $ k $-designs, as defined by using the $ k $-fold channel~\eqref{eq:k-fold_channel} as follows.
An ensemble of unitaries $ \nu $ is a unitary $ k $-design, if $  \Phi_{\nu}^{(k)}(O)
=
\Phi_{\haar}^{(k)}(O) $ holds for \textit{all} $ O\in \bop(\Hsp^{\otimes k}) $,
which implies that all the $ k $th moments of $ \nu $ equal those of the HRU~\cite{DiVincenzo2002,Gross2007,Dankert2009}.
The previous studies~\cite{Roberts2017,Cotler2017,Hunter-Jones2018,Lloyd2018,Lloyd2018full,Zhuang2019} revealed the
fundamental relationship between unitary $ k $-designs and information scrambling.
As discussed below,
on the other hand,
there is a subtle issue about the relationship
between unitary $ k $-designs and Hamiltonian dynamics~\cite{Roberts2017}.

If the $ k $-ETH~\eqref{eq:k-ETH} was satisfied for all observables,
the LTE would become a unitary $ k $-design,
because in that case  Eq.~\eqref{eq:LTE_k-design} holds for all $ O\in \bop(\Hsp^{\otimes k}) $.
In reality, however, there are exceptional observables that do not satisfy Eq.~\eqref{eq:LTE_k-design} (and thus the $ k $-ETH) even approximately.
For example, the projection to an energy eigenstate $ \ket{E_i}\bra{E_i} $ does not satisfy Eq.~\eqref{eq:LTE_k-design}  because of the energy conservation:
\begin{align}
\Phi^{(1)}_\LTE(\ket{E_i}\bra{E_i} )=\ket{E_i}\bra{E_i} \neq\Phi^{(1)}_\haar(\ket{E_i}\bra{E_i} ).
\end{align}
Thus, the LTE is not a unitary $ k $-design.
This can be also understood from the fact that
$ e^{-iHt} $ is diagonal in the energy eigenbasis
and never randomizes the occupations of the energy eigenstates.

The above observation brings us to the notion of PU $ k $-designs.
First, we notice that
in practice we can hardly measure all of the observables on the Hilbert space
in realistic experimental situations of many-body systems,
because the Hilbert space dimension grows exponentially in the system size.
In particular,
it is a common experimental setting~\cite{Trotzky2011,Langen2013,Gross2007,Neill2016,Wendin2017} that one only has the access to (i.e., the ability to perform measurements on)
few-body observables
such as the total magnetization and  local correlation functions,
that is,
one does not observe highly non-local and many-body observables.
Such limitation of accessibility to observables suggests that 
many-body chaotic dynamics can be indistinguishable from the HRU,
only when one looks at the expectation values of the accessible observables.
On the other hand, 
the expectation value of an observable outside the accessible set
can deviate from the HRU behavior,
which implies that  chaotic Hamiltonian dynamics are not full unitary designs.
Thus, we consider unitary $ k $-designs only on the accessible observables,
which we formally define as follows.

\subsection{Definition of PU designs}

We fix an arbitrary  subset of $ \bop(\Hsp^{\otimes k})$
and denote it by $ \sop $,
which is interpreted as the set of accessible operators.
A PU $ k $-design is the ensemble indistinguishable from the HRU
as long as one measures the operators in $ \sop $. 
A PU $ k $-design is then defined as follows.

\begin{Def}[Partial unitary $ k $-design\label{def:PU_design}]
	An ensemble $ \nu $ is a PU design on $ \sop $ if
	\begin{align}
	\Phi_{\nu}^{(k)}(O)=\Phi_{\haar}^{(k)}(O)
	\label{eq:def_PU_design}
	\end{align}
	for all $ O\in\sop $.
\end{Def}

\begin{figure}
	\includegraphics[width=0.8\linewidth]{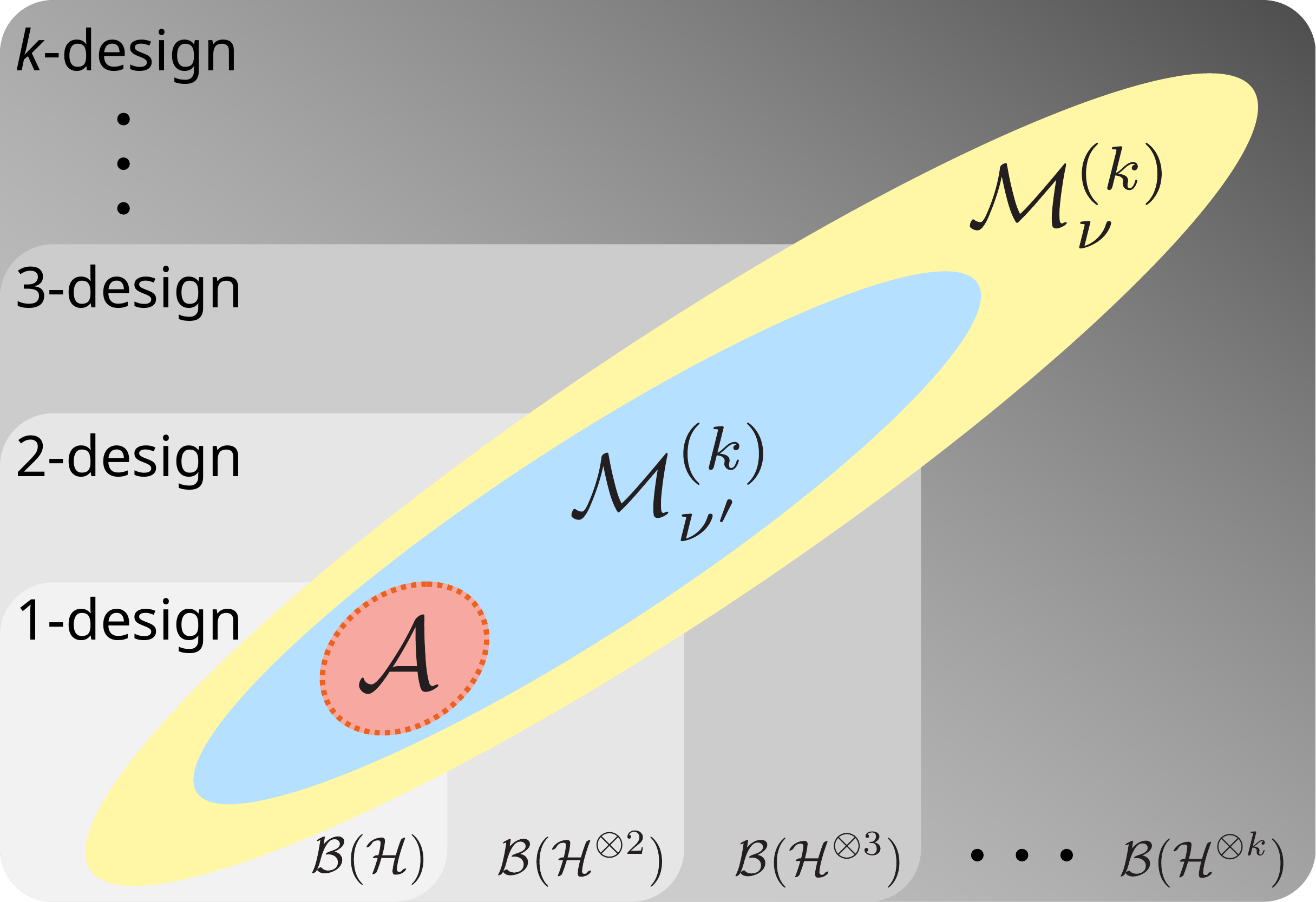}
	\caption{Schematic of PU $ k $-designs.
	$ \mathcal{A} $ is an arbitrary subset of operators, which is interpreted as the set of accessible operators.
	$  \mathcal{M}_{\nu}^{(k)}$ and $ \mathcal{M}_{\nu'}^{(k)} $ are the subsets of  $ \bop(\Hsp^{\otimes k}) $ defined by Eq.~\eqref{eq:scrambled_operators} for given $\nu$ and $\nu'$.
	If $ \mathcal{M}_{\nu}^{(k)} $ includes $ \mathcal{A} $, $ \nu $ is a PU $ k $-design on $ \mathcal{A} $.
	If $ \mathcal{M}_{\nu}^{(k)} $ includes $ \mathcal{M}_{\nu'}^{(k)} $,
	we can say that $ \nu $ is more random than $ \nu' $.
	}
	\label{fig:PU_design}
\end{figure}	

A unitary $ k $-design is a special case of PU $ k $-designs,
where all the operators are accessible:
$\sop=\bop(\Hsp^{\otimes k}) $.
It is also obvious from the definition that
a PU $ k $-design on $ \sop $ is also a PU $ k $-design on $ \sop' $
for any $ \sop'\subset \sop $.
This is a generalization of the fact that
a unitary $ k $-design is also  a unitary ($ k-1 $)-design.

Now fix an ensemble $\nu$.
We define the maximum subset of $ \bop(\Hsp^{\otimes k}) $ such that the ensemble $ \nu $ becomes a PU $ k $-design,
and denote it as
\begin{align}
\mathcal{M}_{\nu}^{(k)}
:= \{O\in \bop(\Hsp^{\otimes k}):
\Phi_{\nu}^{(k)}(O) = \Phi_{\haar}^{(k)}(O)\}.
\label{eq:scrambled_operators}
\end{align}
By using this, the foregoing definition of a PU $ k $-design on $ \sop $ can be rephrased as
\begin{align}
\sop \subset \mathcal{M}_{\nu}^{(k)}
\label{eq:PU-design2}.
\end{align}
See also  Fig.~\ref{fig:PU_design} for a schematic.

For example,
the identity ensemble $ \nu_{\mathrm{id}}:=\{I\} $, where the identity $ I $ appears with probability 1, is the most trivial ensemble.
The maximum subset where $ \nu_{\mathrm{id}} $ becomes a PU $ k $-design is given by a linear combination of the permutation operators
(see Sec.~\ref{sec:PU-design:Example}):
\begin{align}
\mathcal{M}_{\nu_{\mathrm{id}}}^{(k)}
=
\mathrm{span}\left\lbrace
	W_{\pi} :\pi \in S_k
\right\rbrace,
\label{eq:identity_ens_design}
\end{align}
where $  \mathrm{span}\left\{\cdots\right\} $ represents the operator space spanned by $ \{\cdots\} $.
We note that
\begin{align}
\mathcal{M}_{\nu_{\mathrm{id}}}^{(k)}
\subset
\mathcal{M}_{\nu}^{(k)}
\subset
 \mathcal{M}_{\haar}^{(k)}
\end{align}
holds for any ensemble $ \nu $.
$ \mathcal{M}_{\nu}^{(k)}=\mathcal{M}_{\haar}^{(k)} $ is achieved only if $ \nu $ is a unitary $ k $-design.

In general, the subset $ \mathcal{M}_{\nu}^{(k)} $ allows us for quantifying  partial pseudo-randomness of ensemble $ \nu $.
We can say that an ensemble $ \nu $ is more random than an ensemble $ \nu' $, if 
\begin{align}
\mathcal{M}_{\nu'}^{(k)}
\subset 
\mathcal{M}_{\nu}^{(k)}
\label{eq:subset_relation}
\end{align}
holds for all $ k $.
In fact, if an ensemble is indistinguishable from the HRU for a larger set of operators than another ensemble,
the former is regarded as intrinsically more random than the latter
(see also Fig.~\ref{fig:PU_design}).
Specifically, the dimension of $ \mathcal{M}_{\nu}^{(k)} $ as an operator space (i.e., the number of generators of $ \mathcal{M}_{\nu}^{(k)} $) provides quantitative characterization of partial pseudo-randomness of ensemble $ \nu $,
because we have
\begin{align}
\dim(\mathcal{M}_{\nu'}^{(k)})\leq \dim(\mathcal{M}_{\nu}^{(k)}),
\label{eq:dimension_relation}
\end{align}
if Eq.~\eqref{eq:subset_relation} holds.

For the LTE $ \LTE $, $ \mathcal{M}_{\LTE}^{(k)} $ equals the set of operators that satisfy the exact $ k $-ETH~\eqref{eq:k-ETH}.
In other words, the $ k $-ETH is regarded as a specific case of PU $ k $-designs, where pseudo-randomness originates from the random-time sampling of quantum chaotic dynamics.
We show an explicit expression of $ \mathcal{M}_{\LTE}^{(k)} $ in Sec.~\ref{sec:PU-design:Example}.

\subsection{Examples
\label{sec:PU-design:Example}}
We show some special examples of  PU designs that are not full unitary designs.

\subsubsection{Single unitary}
The first example is a trivial case, where the ensemble consists of just a single unitary operator.
We fix a unitary operator $ U\in\uop(\Hsp) $ and denote the ensemble by $ \nu_U:=\{U\} $.
The identity ensemble mentioned above is a special case of $ \nu_{U} $ with $ U=I $.
Obviously, $ \nu_{U} $ is not a unitary $ k $-design for any $ k $.
The maximum subset of $ \bop(\Hsp^{\otimes k}) $ that makes $ \nu_{U} $ a PU $ k $-design is given by the linear combination of the permutation operators:
\begin{align}
\mathcal{M}_{\nu_{U}}^{(k)}
=
\mathrm{span}\left\lbrace
	W_{\pi} :\pi \in S_k
\right\rbrace.
\label{eq:single_ens_design}
\end{align}
We note that $ \dim(\mathcal{M}_{\nu_{U}}^{(k)})=k! $, which does not scale with the system size.
Equation~\eqref{eq:single_ens_design} is proven as follows.
If $ O\in \mathcal{M}_{\nu_{U}}^{(k)}$, we have
\begin{align}
U^{\dag \otimes k} O U^{\otimes k}
=
\Phi_{\haar}^{(k)}(O).
\end{align}
Multiplying this by $U^{\otimes k}  $ from left and by $ U^{\dag \otimes k} $ from  right, we have
\begin{align}
O
=U^{\otimes k}\Phi_{\haar}^{(k)}(O)U^{\dag \otimes k}
=\Phi_{\haar}^{(k)}(O),
\end{align}
where we used the left and right invariance of the HRU.
Equation~\eqref{eq:Haar_k-channel} implies that $ \Phi_{\haar}^{(k)}(O) $ is a linear combination of the permutation operators.
Thus, we obtain 
$  \mathcal{M}_{\nu_{U}}^{(k)} \subset  \mathrm{span}\left\lbrace W_{\pi} :\pi \in S_k \right\rbrace$.
Because
$  \mathcal{M}_{\nu_{U}}^{(k)}\supset \mathrm{span}\left\lbrace W_{\pi} :\pi \in S_k \right\rbrace$
is obvious, we obtain Eq.~\eqref{eq:single_ens_design}.

\subsubsection{Random diagonal-unitaries}
The second example is the ensemble of random diagonal-unitaries~\cite{Nakata2017,Nakata2012,Nakata2014}, which includes the LTE $ \LTE $ as a special case.
We will derive an explicit expression of  the set of operators that satisfy the exact $k$-ETH~\eqref{eq:k-ETH}.

Let $ J:=\{\ket{j}\}_{j\in\{1,\dots, d\}} $ be an orthonormal basis of $ \Hsp $.
A random diagonal-unitary in the basis $ J$ is given by $ U=\sum_{j=1}^de^{i\phi_j}\ket{j}\bra{j} $ with $  \phi_j$ distributing uniformly over $ [0,2\pi] $.
We denote the ensemble of random diagonal-unitaries in the basis $ J$ by $ \RDU_J $.
The maximum subset of $ \bop(\Hsp^{\otimes k}) $ that makes $ \RDU_J $ a PU $ k $-design is given by
\begin{align}
\mathcal{M}_{\RDU_J}^{(k)}
&=
\mathrm{span}\{
W_{\pi}, \ket{\bm{j}}\bra{\bm{j}'}:
\pi\in S_k, \bm{j}\not\sim \bm{j}'
\},
\label{eq:RDU_design}
\end{align}
where $\bm{j}:=(j_1,\dots, j_k) $, $\bm{j}':=(j_1',\dots, j_k') $,
and $ \bm{j}\not\sim \bm{j}'  $ denotes  that $ (j_1,\dots, j_k)$ is not a permutation of $ (j_1',\dots, j_k')  $.

Equation~\eqref{eq:RDU_design} is proven as follows.
The $ k $-fold channel of $ \RDU_J $ is given by
\begin{align}
\Phi_{\RDU_J}^{(k)}(O)
&=
\int_0^{2\pi}\prod_{j=1}^d\frac{d\phi_j}{2\pi}
\sum_{\bm{j},\bm{j}'}e^{-i\sum_i(\phi_{j_i}-\phi_{j_i'})}
\ket{\bm{j}}\braket{\bm{j}|O|\bm{j}'}\bra{\bm{j}'}
\\
&=
\sum_{\bm{j}\sim \bm{j}'} \ket{\bm{j}}\braket{\bm{j}|O|\bm{j}'}\bra{\bm{j}'},
\label{eq:RDU_k-channel}
\end{align}
where $ \bm{j}\sim \bm{j}'  $ denotes that $ (j_1,\dots, j_k)$ is a permutation of $ (j_1',\dots, j_k')  $.
Any operator $ O \in \bop(\Hsp^{\otimes k})$ can be expanded as
\begin{align}
O
=\sum_{\bm{j}\sim \bm{j}'} \ket{\bm{j}}\braket{\bm{j}|O|\bm{j}'}\bra{\bm{j}'}+\sum_{\bm{j}\not\sim \bm{j}'} \ket{\bm{j}}\braket{\bm{j}|O|\bm{j}'}\bra{\bm{j}'}.
\label{eq:decomposition_operator}
\end{align}
From Eq.~\eqref{eq:RDU_k-channel} and Eq.~\eqref{eq:decomposition_operator}, if $ O\in\mathcal{M}_{\RDU_J}^{(k)} $, we have
\begin{align}
O
=\Phi_{\haar}^{(k)}(O)
+\sum_{\bm{j}\not\sim \bm{j}'} \ket{\bm{j}}\braket{\bm{j}|O|\bm{j}'}\bra{\bm{j}'}.
\end{align}
Thus, we obtain 
$  \mathcal{M}_{\RDU_J}^{(k)}\subset\mathrm{span}\{W_{\pi}, \ket{\bm{j}}\bra{\bm{j}'}:\pi\in S_k, \bm{j}\not\sim \bm{j}'\}$.
Because
$   \mathcal{M}_{\RDU_J}^{(k)}\supset\mathrm{span}\{W_{\pi}, \ket{\bm{j}}\bra{\bm{j}'}:\pi\in S_k, \bm{j}\not\sim \bm{j}'\}$
is obvious, we obtain Eq.~\eqref{eq:RDU_design}.

Let us consider the LTE.
As in Sec.~\ref{sec:k-ETH:definition}, we assume that the Hamiltonian satisfies the $ k $th-incommensurate condition.
As mentioned in Sec.~\ref{sec:k-ETH:definition}, the LTE is a diagonal-unitary $ k $-design in the energy eigenbasis $ E:=\{\ket{E_j}\}_{j\in\{1,\dots, d\}} $.
Therefore, from Eq.~\eqref{eq:RDU_design}, we obtain
\begin{align}
\mathcal{M}_{\LTE}^{(k)}
&=\mathcal{M}_{\RDU_E}^{(k)}
\\
&=
\mathrm{span}\{
W_{\pi}, \ket{E(\bm{i})}\bra{E(\bm{i}')}:
\pi\in S_k, \bm{i}\not\sim \bm{i}'
\},
\label{eq:LTE_design}
\end{align}
where $ \ket{E(\bm{i})}:=\ket{E_{i_1}\cdots E_{i_k}} $.
Equation~\eqref{eq:LTE_design} is the main result of this subsection
and completely characterizes the operators that satisfy the exact $ k $-ETH~\eqref{eq:k-ETH}.

\subsubsection{Pauli ensemble}

We next consider the uniform ensemble of the many-body Pauli operators,
which is not a full unitary 2-design but can be a PU 2-design.
We denote the set of the Pauli operators of a single qubit by $ \sPauli:=\{I,\ X,\ Y,\ Z\} $.
The set of the many-body Pauli operators acting on $ N $ qubits consists of $ N $ tensor products of elements of $\sPauli$,
denoted as $ \sPauli_N:=\sPauli^{\otimes N} $.
The total number of elements in $ \sPauli_N $ is $ 4^N=d^2 $.

We denote the uniformly distributed ensemble over $\sPauli_N  $ by $ \Pauli $.
It is known that $ \Pauli $ is a unitary 1-design~\cite{Low2010}, and thus
\begin{align}
\mathcal{M}_{\Pauli}^{(1)}=\bop(\Hsp).
\end{align}
On the other hand, $ \Pauli $ is not a unitary 2-design~\cite{Low2010}.
In fact,
if $ N=1 $ and $ O=X^{\otimes 2} $, we have
\begin{align}
\Phi^{(2)}_{\Pauli}(X^{\otimes 2})
&=
\frac{1}{4}\left(
I^{\otimes 2}X^{\otimes 2}I^{\otimes 2}+X^{\otimes 2}X^{\otimes 2}X^{\otimes 2}
\right.
\nonumber\\
&
\qquad
\left.
+Y^{\otimes 2}X^{\otimes 2}Y^{\otimes 2}+Z^{\otimes 2}X^{\otimes 2}Z^{\otimes 2}
\right)
\\
&=
X^{\otimes 2},
\end{align}
which does not equal
$ \Phi^{(2)}_{\haar}(X^{\otimes 2})
=-\frac{1}{3}I^{\otimes 2}
+\frac{2}{3}S
$
with $S$ being the swap operator introduced in Eq.~\eqref{eq:swap}.

On the other hand,
the maximum set of operators such that $ \Pauli$ become a PU 2-design is given by a linear combination of Pauli operators and the swap operator:
\begin{align}
\mathcal{M}_{\Pauli}^{(2)}
=
\mathrm{span}
\{W_{\pi}, P_l\otimes P_m\in \sPauli_N^{\otimes 2} :\pi \in S_2,  l\neq m \}.
\label{eq:Pauli_PU-design}
\end{align}
We note that the number of the generators of $ \mathcal{M}_{\Pauli}^{(2)} $ is $ d^4-d^2+2 $ (i.e.,  $ \dim(\mathcal{M}_{\Pauli}^{(2)})=d^4-d^2+2 $),
which is close to the maximum value $ \dim(\bop(\Hsp^{\otimes 2}))=d^4 $.
This is contrastive to the fact that $ \mathcal{M}_{\nu_{U}}^{(2)}= \mathrm{span}\{ I, S\} $ for a single unitary has only two generators for $k=2$.
In other words, $ \Pauli $ is very close to a full unitary 2-design in the sense of  the argument around Eq.~\eqref{eq:dimension_relation}.

Equation~\eqref{eq:Pauli_PU-design} is proven as follows.
First, any $P_l \in \sPauli_N$ with $P_l \neq I$ commutes with a half of the elements of $\sPauli_N$ and anti-commutes with the other half.
Then, we introduce a function $ F(l,m) $ for any $P_l, P_m \in \sPauli_N$ by
\begin{align}
F(l,m)
:=
\begin{dcases}
0 & \text{for}\ P_lP_m=P_lP_m\\
1 & \text{for}\ P_lP_m=-P_mP_l.
\end{dcases}
\end{align}
Because $ \sPauli_N $ forms an orthonormal basis of $ \bop(\Hsp) $
with the Hilbert-Schmidt inner product $  \braket{A|B}:=\Dsh^{-1}\tr{A^\dag B}$,
any operator $ O \in \bop(\Hsp^{\otimes 2})$ can be expanded as
$ O=\sum_{l,m}c_{lm}P_l\otimes P_m $,
where $ c_{lm}:=\braket{P_l\otimes P_m|O}$.
Thus,
the 2-fold channel of $ \Pauli $ is written as
\begin{align}
\Phi_{\Pauli }^{(2)}(O)
&=
\frac{1}{d^2}
\sum_{j,l,m}c_{lm}(P_jP_lP_j)\otimes (P_jP_mP_j)
\\
&=
\frac{1}{d^2}
\sum_{j,l,m}c_{lm}(-1)^{F(j,l)+F(j,m)} P_l\otimes P_m
\\
&=
\sum_{l}c_{ll}P_l\otimes P_l,
\label{eq:2-Pauli_channel}
\end{align}
where we used
$ d^{-2}\sum_{j}(-1)^{F(j,l)+F(j,m)} =\delta_{lm} $
to obtain the final line.
Comparing  Eq.~\eqref{eq:2-Haar_channel} and Eq.~\eqref{eq:2-Pauli_channel},
we obtain
\begin{align}
c_{ll}
&=
\begin{dcases}
\frac{\tr{O}}{d^2}
\quad & (P_l=I)
\\
\frac{d\tr{SO}-\tr{O}}{d^2(d^2-1)}
\quad & (P_l\neq I),
\end{dcases}
\end{align}
as a necessary and sufficient condition for
$ \Phi_{\Pauli}^{(2)}(O)=\Phi_{\haar}^{(2)}(O) $.
In other words,
$ O $ is written as
\begin{align}
O
&=
c_II^{\otimes 2}+c_SS+\sum_{l\neq m}c_{lm}P_l\otimes P_m,
\label{eq:Pauli_PU-design_condition1}
\end{align}
where
\begin{align}
c_I=\frac{d\tr{O}-\tr{SO}}{d(d^2-1)},\  c_S=\frac{d\tr{SO}-\tr{O}}{d(d^2-1)}.
\label{eq:Pauli_PU-design_condition2}
\end{align}
Therefore,
we obtain Eq.~\eqref{eq:Pauli_PU-design}.

\section{Summary and Discussion
	\label{sec:Discussion}}
In this paper, we have proposed the $ k $-ETH~\eqref{eq:k-ETH} to characterize complexity of chaotic dynamics in quantum many-body systems.
A positive integer $ k $ stands for the $k$th moment of dynamics and is regarded as the order of complexity.
While the 1-ETH is the conventional ETH, the $ k $-ETH ($ k\geq 2 $) gives deeper characterization of quantum chaos beyond thermalization.
In particular, the $ k $-ETH ($ k \geq 2 $) gives a non-negligible contribution to the $ k $-REE~\eqref{eq:thermal_k-REE}, which corresponds to the Page correction for individual energy eigenstates.
By using numerical exact diagonalization,
we have confirmed that the 2-ETH is approximately true with respect to a proper norm~\eqref{eq:2-ETH_indicator}, except for the integrable case (Table~\ref{table:summary}).

Our formulation of the $k$-ETH is intrinsically related to unitary $k$-designs.
To further elaborate this relationship, we have proposed a new information-theoretic concept of quantum pseudo-randomness: PU $ k $-designs (Definition~\ref{def:PU_design}).
The $ k $-ETH is a specific example of PU $ k $-designs
of the LTE.
As anticipated from the connection between information scrambling and unitary $k$-designs~\cite{Roberts2017,Cotler2017}, the exact $ k $-ETH implies that
the long-time average of the $ 2k $-point OTOC
equals the exact HRU average~\eqref{eq:exact_scrambling}.
However, we have pointed out that, as for the OTOC,
the difference between the 1-ETH and the higher-order ETH is smaller than the finite-size effect in quantum many-body systems.

Our theory would be regarded as a first step for understanding why random dynamics can well approximate chaotic quantum dynamics,  nevertheless the Hamiltonian in itself does not have randomness.
An immediate open question is whether the $k$-ETH with $k \geq 3$ is true in many-body systems, for which we need more comprehensive numerical studies with various  non-integrable models. 
We note that the Page curve for the 3-REE has been numerically investigated in Ref.~\cite{Lu2019}.
Also, the validation of the weak $k$-ETH in integrable models is an open issue, as well as the possibility of many-body scar~\cite{Bernien2017,Turner2018} of the higher-order ETH.

Toward more profound understanding of quantum chaos and many-body complexity, there are numerous  issues to investigate in the future.
For example,  it would be important to bridge the higher-order ETH with the early time scrambling behavior beyond our late-time characterization.
In fact, a recent study~\cite{Murthy2019} rederived
the Maldacena-Shenker-Stanford~(MSS) bound on OTOCs~\cite{Maldacena2016JHEP}
from the ETH in the form of Srednicki's ansatz~\cite{Srednicki1994}.
To get a unified view on their approach and ours is an interesting future direction.

Recently,
the circuit complexity,
a more direct measure of quantum complexity,
has been studied in the context of black hole geometry~\cite{Susskind2016,Brown2017,Brown2018}.
In the circuit model,
thermalization and information scrambling are accompanied by 
growth of the circuit complexity,
and therefore it is interesting to investigate the circuit-complexity aspect of
our PU design formulation.
It is known that
the circuit complexity of a unitary $ k $-design is linearly lower-bounded by
the number of qubits~\cite{Roberts2017}.
However,
a direct connection between the circuit complexity and PU $ k $-designs
is not known so far.

Nowadays,
chaotic many-body dynamics have been studied extensively
in experiments with isolated quantum systems, such as ultracold atoms~\cite{Trotzky2011,Langen2013,Gross2017} and superconducting qubits~\cite{Neill2016,Wendin2017}.
The 2-REE has been observed in a recent experiment with ultracold atoms~\cite{Islam2015,Kaufman2016}.
There are also several proposals of measuring  OTOCs~\cite{Swingle2016,Zhu2016,Halpern2016}, and OTOCs have been indeed measured with trapped ions~\cite{Garttner2017}.
Furthermore, unitary designs are experimentally realized with nuclear magnetic resonance systems~\cite{Ryan2009,Li2019}.
In light of the development of such state-of-the-art technologies,
the experimental study of higher-order complexity and quantum pseudo-randomness
is an interesting future issue,
because these experimental setups would reach much larger-scale quantum many-body systems beyond the numerically accessible scale.

\begin{acknowledgments}
	K.K. is supported by JSPS KAKENHI Grant No.~JP17J06875.
	E.I. and T.S. are supported by JSPS KAKENHI Grant No.~JP16H02211.
	E.I. is supported by JSPS KAKENHI Grant No.~JP15K20944 and JP19K14609.
	T.S. is supported by JSPS KAKENHI Grant No.~JP19H05796.
\end{acknowledgments}

\appendix

\section{Proofs
\label{sec:Proof}}
In this appendix,
we prove several results of the main text.

\subsection{Derivation of Eq.~\eqref{eq:k-ETH}
\label{sec:Proof:k-ETH}}

The $ k $-fold channel of  the LTE is written in terms of the energy eigenbasis as
\begin{align}
\Phi_{\LTE}^{(k)}(O)
=
\sum_{\bm{i},\bm{j}}
S(\bm{i},\bm{j})
\ket{E(\bm{i})}\bra{E(\bm{j})},
\end{align}
where
\begin{align}
S(\bm{i},\bm{j}):&=
\lim_{T\to\infty}\frac{1}{T}\int_0^Tdt
\exp\left[
it\left(
\sum_{m=1}^k (E_{i_m}-E_{j_m})
\right)
\right]
\nonumber\\
&\qquad\times\braket{E(\bm{i})|O|E(\bm{j})}
\label{eq:LTE_k-channel_matrix}
\end{align}
with $ \ket{E(\bm{i})}:=\ket{E_{i_1}\cdots E_{i_k}} $.
If the spectrum of $ H $ is $ k $th-incommensurate, we have
\begin{align}
&\lim_{T\to\infty}\frac{1}{T}\int_0^Tdt
\exp\left[
it\left(
\sum_{m=1}^k (E_{i_m}-E_{j_m})
\right)
\right]
\nonumber\\
&=
\begin{dcases}
1 &(\bm{i}\sim\bm{j})\\
0 &(\bm{i}\not\sim\bm{j}),
\end{dcases}
\end{align}
where $ \bm{i}\sim \bm{j} $ ($ \bm{i}\not\sim \bm{j} $) denotes that $ (i_1,\dots, i_k)$ is (not) a permutation of $ (j_1,\dots, j_k)  $.
Substituting this into Eq.~\eqref{eq:LTE_k-channel_matrix},
we obtain
\begin{align}
\Phi_{\LTE}^{(k)}(O)=
\sum_{\bm{i}\sim\bm{j}}
\ket{E(\bm{i})}
\braket{E(\bm{i})|O|E(\bm{j})}
\bra{E(\bm{j})}.
\label{eq:LTE_k-channel}
\end{align}
By comparing the matrix elements
of Eq.~\eqref{eq:Haar_k-channel}
and Eq.~\eqref{eq:LTE_k-channel},
we have
\begin{align}
&\braket{E_{i_1}\cdots E_{i_k}|O|E_{i_{\sigma(1)}}\cdots E_{i_{\sigma(k)}}}
\nonumber\\
&=
\sum_{\pi,\tau\in S_k}\Wg_{\pi,\tau}(\Dsh)\tr{W_\tau O}
\braket{E_{i_1}\cdots E_{i_k}|W_\pi |E_{i_{\sigma(1)}}\cdots E_{i_{\sigma(k)}}}
\\
&=
\sum_{\pi,\tau\in S_k}
\delta_{\pi\sigma}(\bm{i})
\Wg_{\pi,\tau}(\Dsh)\tr{W_\tau O}
\end{align}
for any $ \sigma \in S_k $.
All other matrix elements of $ \Phi_{\haar}^{(k)}(O) $ and $ \Phi_{\LTE}^{(k)}(O) $ are zero.
Therefore, the $ k $-ETH~\eqref{eq:k-ETH} is a necessary and sufficient condition for the $k$th-order quantum ergodicity~\eqref{eq:LTE_k-design}.

\subsection{Proof of Eq.~\eqref{eq:k-ETH_product_asymptotic}
\label{sec:Proof:product}}
For large $ \Dsh $, the Weingarten matrix $ \Wg_{\pi,\tau}(d) $ scales as follows~\cite{Colins2006}:
\begin{align}
\Wg_{\pi,\tau}(d)=\mathcal{O}(\Dsh^{-2k+c(\pi\tau)}),
\label{eq:Weingarten_asymptotic}
\end{align}
where the dominant term is given by $ \tau=\pi^{-1} $:
\begin{align}
\Wg_{\pi,\pi^{-1}}(d)=\Dsh^{-k}+\mathcal{O}(\Dsh^{-k+2}).
\label{eq:Weingarten_identity}
\end{align}
From the assumption that $  \braket{A^l}_{\mathrm{mc}}=\mathcal{O}(1) $, we have
\begin{align}
\prod_{m=1}^{c(\tau)}\tr{A^{l_m(\tau)}}
=
\Dsh^{c(\tau)}\prod_{m=1}^{c(\tau)}\braket{A^{l_m(\tau)}}_{\mathrm{mc}}
=
\mathcal{O}(\Dsh^{c(\tau)}).
\end{align}
From this and Eq.~\eqref{eq:Weingarten_asymptotic},
we have
\begin{align}
\Wg_{\pi,\tau}(d)\prod_{m=1}^{c(\tau)}\tr{A^{l_m(\tau)}}
&=
\mathcal{O}(\Dsh^{-2k+c(\pi\tau)+c(\tau)})
\\
&=
\mathcal{O}(\Dsh^{-k+c(\pi)}),
\end{align}
where we used $ c(\pi\tau)\leq k+c(\pi)-c(\tau)$~\cite{Goldstein2014}.
Because of the fact that $ c(\pi)=k $ holds if and only if $ \pi $ is the identity, we obtain
\begin{align}
\sum_{\tau\in S_k}\Wg_{\pi,\tau}(d)\prod_{m=1}^{c(\tau)}\tr{A^{l_m(\tau)}}
=
\begin{dcases}
 \mathcal{O}(1) &\text{for}\ \pi =I\\
 \mathcal{O}(\Dsh^{-1}) &\text{for}\ \pi \neq I.
\end{dcases}
\end{align}
Thus,  the right-hand side of Eq.~\eqref{eq:k-ETH_product} is $  \mathcal{O}(\Dsh^{-1})  $ unless $ \delta_{\sigma}(\bm{i})=1 $, or equivalently  $ i_1=i_{\sigma(1)},\dots, i_k=i_{\sigma(k)} $.
If $ i_1=i_{\sigma(1)},\dots, i_k=i_{\sigma(k)} $, the leading term of the right-hand side of Eq.~\eqref{eq:k-ETH_product} is given by $ \pi=\tau=I $,
which asymptotically equals $ \Dsh^{-2k+k}\cdot \Dsh^{k}\braket{A}_{\mathrm{mc}}^k=\braket{A}_{\mathrm{mc}}^k $.
This proves Eq.~\eqref{eq:k-ETH_product_asymptotic}.

\subsection{Proof of Eq.~\eqref{eq:typical_error}
\label{sec:Proof:typical_error}}

Let $ \{A_1,\dots,A_{\Dsh^2}\} $ be an orthonormal basis for the operator space $ \bop(\Hsh) $ with respect to the Hilbert-Schmidt inner product.
Then, 
any operator $ O\in\bop(\Hsh^{\otimes k}) $ can be expanded as
\begin{align}
O=\sum_{i_1,\dots, i_k}c_{i_1,\dots, i_k}A_{i_1}\otimes\cdots \otimes  A_{i_k},
\label{eq:operator_expansion}
\end{align}
where $ c_{i_1,\dots, i_k}:=\braket{A_{i_1}\otimes\cdots \otimes  A_{i_k}|O} $.
The normalization $ \braket{O|O}=1 $ is equivalent to 
$\sum_{i_1,\dots, i_k}\abs{c_{i_1,\dots, i_k}}^2=1 $.
This leads to a one-to-one correspondence between a normalized operator $O$ and a normalized vector $ \ket{O} $ (regarded as a quantum state),
which is referred to as the channel-state duality~\cite{Choi1975,Jamiolkowski1972} (see also Appendix~\ref{sec:OSEE}).
Also, $ \{c_{i_1,\dots, i_k}\}$ are identified with a point of the ($ 2\Dsh^{2k}-1 $)-dimensional unit hypersphere.

In the following,
we consider a random operator $ O $ expressed as Eq.~\eqref{eq:operator_expansion},
where $\{c_{i_1,\dots, i_k}\}$ are distributed according to the uniform ensemble on the ($ 2\Dsh^{2k}-1 $)-dimensional unit hypersphere.
We denote the average over this ensemble by $ \braket{\cdots} $.
Equation~\eqref{eq:typical_error} is now rigorously stated as follows.
Let $ O $ be  a random operator of $\bop(\Hsh^{\otimes k})  $
and consider the $ k $-ETH indicator $ I_k(O) $ 
defined in Eq.~\eqref{eq:2-ETH_indicator}.
Then,
\begin{align}
\mathrm{Prob}\left[
I_k(O)\geq \Dsh^{-k/2+\delta}
\right]
\leq 2(k!)\Dsh^{2k}\exp\left(
-\frac{2\Dsh^{-2\delta}}{36\pi^3}
\right)
\label{eq:approximate_k-ETH_typical}
\end{align}
holds for any $ \delta>0 $.
Because the right-hand side of Eq.~\eqref{eq:approximate_k-ETH_typical}
decays double exponentially in the system size,
$ I_k(O)=\mathcal{O}(\Dsh^{-k/2+\delta}) $ (i.e., the approximate $ k $-ETH) holds for almost all operators.

The proof of the above statement is given in the same manner as in Ref.~\cite{Reimann2015}.
The key of the proof is the Levy's lemma~\cite{Ledoux2001} stating that
for any $\varepsilon > 0$ and any Lipschitz continuous function $f$ with the Lipschitz constant $\eta$, 
\begin{align}
\mathrm{Prob}\left[
\abs{f(x)-\braket{f(x)}}\geq \epsilon
\right]
\leq 2\exp\left(
-\frac{\epsilon^2 n}{9\pi^3\eta^2}
\right)
\end{align}
holds if $x$ is uniformly randomly sampled from the $ (n-1) $-dimensional unit hypersphere.
We apply the Levy's lemma to $ \Delta(i_1,\dots, i_k;\sigma) $ defined in Eq.~\eqref{eq:k-ETH_error}.
Note that the average of $ \Delta(i_1,\dots, i_k;\sigma) $ is exactly zero,
and the Lipschitz constant is bounded as $ \eta \leq 2\Dsh^{k/2} $.
The Levy's lemma now implies that
\begin{align}
\mathrm{Prob}\left[
\abs{\Delta(i_1,\dots, i_k;\sigma)}\geq \epsilon
\right]
\leq 2\exp\left(
-\frac{2\epsilon^2 \Dsh^k}{36\pi^3}
\right).
\end{align}
Therefore,
we obtain
\begin{align}
\mathrm{Prob}\left[
I_k(O)\geq \epsilon
\right]
\leq 2(k!)\Dsh^{2k}\exp\left(
-\frac{2\epsilon^2 \Dsh^k}{36\pi^3}
\right),
\end{align}
and complete the proof by setting $ \epsilon = \Dsh^{-k/2+\delta}$.

\subsection{Proof of Eq.~\eqref{eq:exact_scrambling}
\label{sec:Proof:exact_scrambling}}

\begin{figure}
	\centering
	\includegraphics[width=0.6\linewidth]{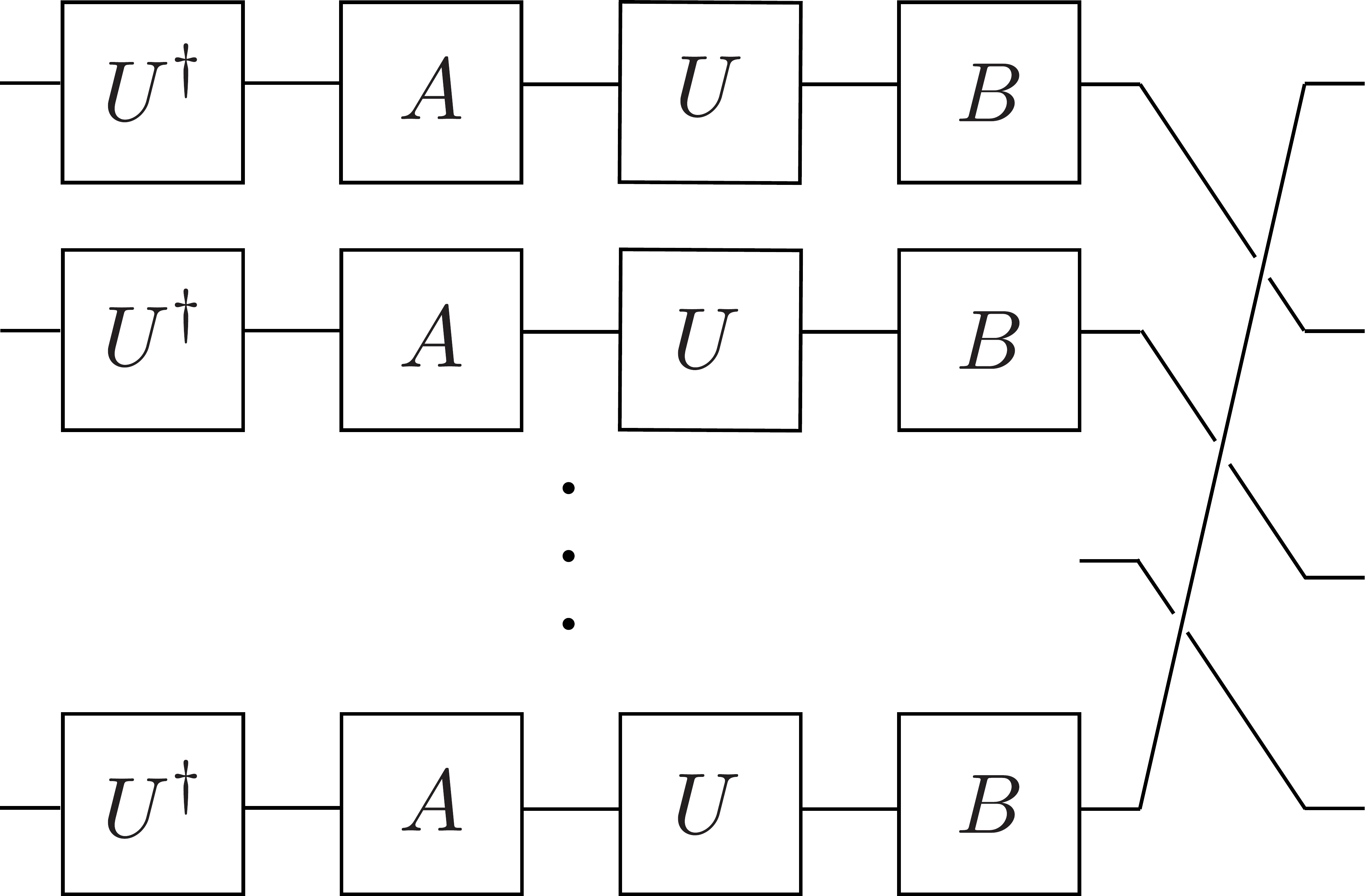}
	\caption{\label{fig:OTOC}
		Graphical representation of the right-hand side of Eq.~\eqref{eq:OTOC_k-channel},
		where we impose the periodic boundary condition.
	}
\end{figure}

We represent the $ 2k $-point OTOC by the $ k $-fold channel.
The trace of the product of $ k $ operators
can be written as
\begin{align}
\tr{A_1\cdots A_k}
=\tr{W_{\mathrm{cyc}}(A_1\otimes \cdots \otimes A_k)}, 
\end{align}
where $ W_{\mathrm{cyc}} $ represents the cyclic permutation operator
on $ \Hsh^{\otimes k} $.
By using the above relation,
we have
\begin{align}
&\ave{\nu}{
\braket{(UAU^\dag)B\cdots (UAU^\dag)B}_{\mathrm{mc}}}
\nonumber\\
&\qquad=
\frac{1}{\Dsh}\tr{\Phi_\nu^{(k)}(A^{\otimes k})B^{\otimes k}W_{\mathrm{cyc}}},
\label{eq:OTOC_k-channel}
\end{align}
which is graphically represented in Fig.~\ref{fig:OTOC}.
Because the $ k $-ETH for $ A^{\otimes k} $ implies that
$ \Phi_\LTE^{(k)}(A^{\otimes k})=\Phi_\haar^{(k)}(A^{\otimes k}) $,
we obtain Eq.~\eqref{eq:exact_scrambling}.

\subsection{Proof of Eq.~\eqref{eq:approximate_scrambling}
\label{sec:Proof:approximate_scrambling}}
From Eq.~\eqref{eq:LTE_k-channel}
and Eq.~\eqref{eq:OTOC_k-channel},
the long-time average of the $ 2k $-point OTOC is represented as
\begin{align}
\overline{F}_{A,B}^{(2k)}
&=
\frac{1}{d}\sum_{\bm{i}\sim\bm{j}}
A_{i_1j_1}\cdots A_{i_kj_k}B_{j_1i_2}\cdots B_{j_ki_1}
\label{eq:2k-OTOC_LTE}.
\end{align}
When $ A_{ij}= \mathcal{O}(\Dsh^{-1/2}) $ and $ B_{ij} = \mathcal{O}(\Dsh^{-1/2}) $, we have
\begin{align}
\overline{F}_{A,B}^{(2k)}
=\mathcal{O}(\Dsh^{-1}) .
\label{eq:2k-OTOC_LTE_traceless}
\end{align}

\section{$ k $-ETH with symmetry
	\label{sec:Symmetry}}

In this appendix, we generalize Eq.~\eqref{eq:k-ETH} in the presence of a unitary symmetry or an anti-unitary symmetry (class AI or AII in the Dyson's classification~\cite{Dyson1962}).
The former simply gives the same form as Eq.~\eqref{eq:k-ETH} for each symmetry sector, as is the case for the 1-ETH~\cite{DAlessio2015}.
The latter, on the other hand, leads to a different form depending on the symmetry class for $k \geq 2$, while the 1-ETH takes the same form for these symmetry classes.

\subsection{Unitary symmetry}
We first briefly note  the case that the Hamiltonian has a unitary symmetry described by a group $ G $.
Let $ \{V(g): g\in G\} $ be a unitary representation of $ G $ on $ \Hsh $.
We can decompose it to irreducible representations as
\begin{align}
\Hsh = \bigoplus_{\lambda} \mathcal{M}_{\lambda}\otimes \mathcal{N}_{\lambda},
\end{align}
where $ \lambda $ is a label of an irreducible representation,
$ \mathcal{M}_{\lambda} $ is the representation space of $ \lambda $,
and $ \mathcal{N}_{\lambda} $ describes the multiplicities of $ \lambda $.
Correspondingly, $ V $ is decomposed as
\begin{align}
V(g)=\bigoplus_{\lambda} V_{\lambda}(g)\otimes I_{\mathcal{N}_\lambda},
\end{align}
where $ I_{\mathcal{N}_\lambda} $ is the identity operator on $ \mathcal{N}_\lambda $.
We assume that the Hamiltonian $ H  $ has the $ G $-symmetry, and commutes with $ V(g) $ for all $ g\in G $.
Because of the Schur's lemma~\cite{Hayashi2017}, the Hamiltonian is decomposed as 
\begin{align}
H=\bigoplus_{\lambda} I_{\mathcal{M}_\lambda}\otimes H_{\lambda},
\end{align}
that is, the Hamiltonian takes a block-diagonal form with each block corresponding to the sector of the irreducible representation (the symmetry sector).
Because the Hamiltonian acts on each sector independently, we focus on a single sector with Hamiltonian $H_\lambda$.
We then obtain the $ k $-ETH for $ H_{\lambda} $
in the same manner as Eq.~\eqref{eq:k-ETH} in Sec.~\ref{sec:k-ETH:definition}:
\begin{align}
&\braket{E_{i_1}^{\lambda}\dots E_{i_k}^{\lambda}|O|E_{i_{\sigma(1)}}^{\lambda}\dots E_{i_{\sigma(k)}}^{\lambda}}
\nonumber
\\
&\qquad=
\sum_{\pi,\tau \in S_k}
\delta_{\pi\sigma}(\bm{i})\Wg_{\pi,\tau}(n_{\lambda})\tr{W_{\tau}O},
\end{align}
where $ O\in \bop(\mathcal{N}_{\lambda}^{\otimes k}) $,
$ \ket{E_i^\lambda} $ is an eigenstate of  $ H_{\lambda} $,
and $ n_{\lambda} $ is the dimension of $ \mathcal{N}_{\lambda} $.
This is just a straightforward extension of the well-established fact that the 1-ETH holds within each symmetry sector~\cite{DAlessio2015}.

\subsection{Anti-unitary symmetry}
We next consider the case that the Hamiltonian has an anti-unitary symmetry.
We here restrict ourselves to class AI and AII, where the Hamiltonian only has an anti-unitary inversion symmetry such as time-reversal symmetry.
We denote the corresponding anti-unitary operator
that commutes with the Hamiltonian by $ T $,
which satisfies $T^2 = 1$ (class AI) or $T^2 = -1$ (class AII).

\subsubsection{Class AI}
We assume that $ H $ does not have any degeneracy and its eigenstates satisfy $ T\ket{E_i}=\ket{E_i} $.
If $H$ is in class AI, the unitary operator $e^{-iHt}$ is in $\mathrm{U}(d)/\mathrm{O}(d) $, where U(d) is the unitary group and $ \mathrm{O}(d) $ is the orthogonal group.
The uniform ensemble over $ \mathrm{U}(d)/\mathrm{O}(d) $ is the circular orthogonal ensemble (COE).
We can then derive the $ k $-ETH of class AI by using the COE instead of the HRU (CUE), which gives
\begin{align}
&\braket{E_{i_1}\dots E_{i_k}|O|E_{i_{\sigma(1)}}\dots E_{i_{\sigma(k)}}}
\nonumber
\\
&=
\sum_{\tau \in S_{2k}}
\Wg^{\mathrm{AI}}(\tau, \Dsh)
\braket{E_{i_1}\dots E_{i_k}|
	O^{\mathrm{AI}}_{\tau}
	|E_{i_{\sigma(1)}}\dots E_{i_{\sigma(k)}}}.
\label{eq:k-ETH_AI}
\end{align}
Here,  $ \Wg^{\mathrm{AI}}(\tau, \Dsh) $ is 
the Weingarten function of COE~\cite{Matsumoto2012,Matsumoto2013},
and
\begin{align}
O^{\mathrm{AI}}_{\tau}:=
\mathrm{tr}_2[W^{\top}_{\pi_\ast\tau\pi_\ast^{-1}}(I\otimes O^{\top})],
\end{align}
where
$ \top $ denotes the transpose,
$ \mathrm{tr}_2 $ denotes the partial trace over the second Hilbert space of $ \Hsh^{\otimes k}\otimes \Hsh^{\otimes k} $,
and
\begin{align}
\pi_\ast:=
\begin{pmatrix}
1 & \cdots k & k+1 &\cdots &2k\\
2 & \cdots 2k & 1 & \cdots  &2k-1
\end{pmatrix}
.
\end{align}
For $ k=1 $, Eq.~\eqref{eq:k-ETH_AI} reduces to the conventional ETH $ \braket{E_i|O|E_i}=\braket{O}_{\mathrm{mc}} $.

We show the derivation of Eq.~\eqref{eq:k-ETH_AI}.
The $ k $-fold channel of the COE is given by~\cite{Matsumoto2012,Matsumoto2013}
\begin{align}
\Phi_{\COE}^{(k)}(O)
=
\sum_{\sigma\in S_{2k}}\Wg^{\mathrm{AI}}(\tau, d)O^{\mathrm{AI}}_{\tau}.
\label{eq:COE_k-channel}
\end{align}
In the same manner as the derivation of the $ k $-ETH of class A,
we compare Eq.~\eqref{eq:LTE_k-channel}
and Eq.~\eqref{eq:COE_k-channel},
and obtain Eq.~\eqref{eq:k-ETH_AI}.

\subsubsection{Class AII}
For a Hamiltonian $H$ in class AII, it always has degeneracies called the Kramers degeneracies. 
Here, we assume that $ H $ have no other degeneracies than the Kramers degeneracies.
We denote an eigenstate of $H$ by $ \ket{E_i, a} $ ($ i=1,\dots,\Dsh $),
where $ a=0,1 $ is the index of the Kramers degeneracy.
We assume that $ T\ket{E_i, 0}=-\ket{E_i, 1}$ and $ T\ket{E_i, 1}=\ket{E_i, 0} $ without loss of generality.

If $H$ is in class AII, the unitary operator $e^{-iHt}$ is in  $ \mathrm{U}(2d)/\mathrm{Sp}(2d) $, where $ \mathrm{Sp}(2d) $ is the symplectic group.
The uniform ensemble over $\mathrm{U}(2d)/\mathrm{Sp}(2d) $ is the circular symplectic ensemble (CSE).
We can derive the $ k $-ETH of class AII by using the CSE instead of the HRU, which gives
\begin{align}
&\braket{E_{i_1}a_1\dots E_{i_k}a_k|O|E_{i_{\sigma(1)}}b_{1}\dots E_{i_{\sigma(k)}}b_{k}}
\nonumber
\\
&=
\sum_{\tau \in S_{2k}}
\Wg^{\mathrm{AII}}(\tau, \Dsh)
\nonumber\\
&\qquad
\times  \braket{E_{i_1}a_1\dots E_{i_k}a_k|
	O^{\mathrm{AII}}_{\tau}
	|E_{i_{\sigma(1)}}b_{1}\dots E_{i_{\sigma(k)}}b_{k}}.
\label{eq:k-ETH_AII}
\end{align}
Here, $ \Wg^{\mathrm{AII}}(\tau, \Dsh) $ is  the Weingarten function of the CSE~\cite{Matsumoto2013}, and
\begin{align}
O^{\mathrm{AII}}_{\tau}:=
\mathrm{tr}_2[W^{\top}_{\pi_\ast\tau\pi_\ast^{-1}}(I_{\Dsh}\otimes O^{\mathrm
D})],
\end{align}
where
$ O^{\mathrm{D}}$ is the dual operator of $ O $ defined as
\begin{align}
O^{\mathrm{D}}:=J^{\otimes k}O^{\top} J^{\top\otimes k}
\end{align}
with
\begin{align}
J:=\begin{pmatrix}
0_{d} & I_{d}\\
-I_{d} & 0_{d}
\end{pmatrix}
.
\end{align}
For $ k=1 $,  Eq.~\eqref{eq:k-ETH_AII} again reduces to the conventional ETH
$ \braket{E_i|O|E_i}=\braket{O}_{\mathrm{mc}} $.
The derivation of Eq.~\eqref{eq:k-ETH_AII} is the same as that of Eq.~\eqref{eq:k-ETH_AI} by using
\begin{align}
\Phi_{\CSE}^{(k)}(O)
=
\sum_{\sigma\in S_{2k}}\Wg^{\mathrm{AII}}(\tau, d)O^{\mathrm{AII}}_{\tau}.
\label{eq:CSE_k-channel}
\end{align}

\section{Operator space entanglement entropy
\label{sec:OSEE}}
We consider the OSEE, which is an analog of the entanglement entropy in the operator space~\cite{Zanardi2001,Prosen2007}.
Specifically, we prove that the $k$th \renyi OSEE ($k$-OSEE) of the LTE follows the Page curve if the $ k $-ETH holds.
This is a natural extension of our result of the entanglement entropy in Sec~\ref{sec:Entanglement}.

\subsection{Definition of the OSEE}
\begin{figure}
	\centering
	\includegraphics[width=0.6\linewidth]{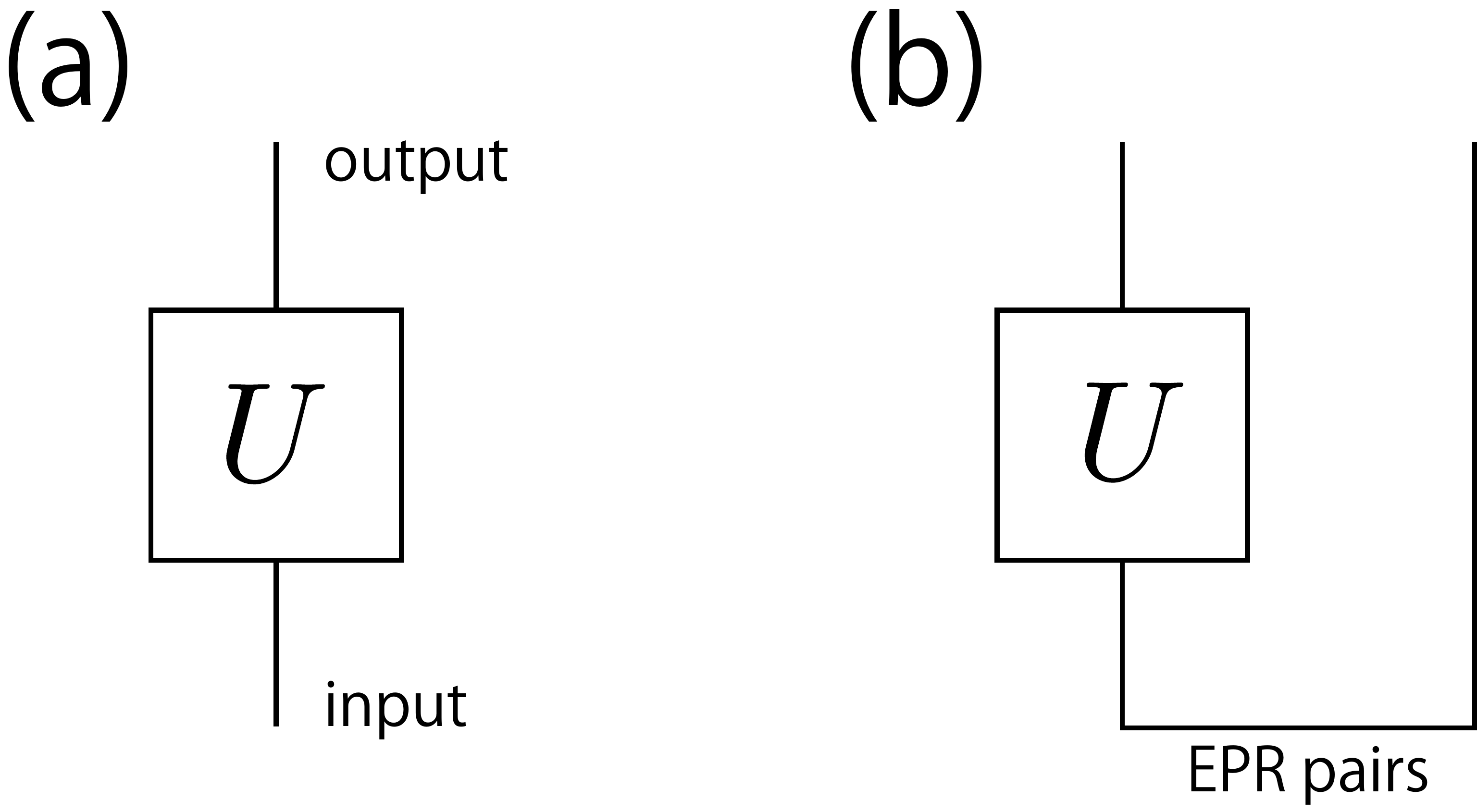}
	\caption{\label{fig:channel-state_duality}
		Graphical representation of the channel-state duality of a unitary operator $ U $.
		(a) The channel representation of $ U $, which has an input leg and an output leg.
		Each leg corresponds to a single Hilbert space $ \Hsp $.
		(b) The corresponding state representation of $ U $.
		By bending the input leg, we treat $ U $ as a state of $ \Hsp^{\otimes 2} $, written as $\ket{U}$.
		The bent line corresponds to $ N $ EPR pairs.
	}
\end{figure}

To formulate the OSEE, we use the channel-state duality (the Choi-Jamio\l{}kowski isomorphism),
which maps a quantum channel (unitary operation) to a quantum state~\cite{Choi1975,Jamiolkowski1972}.
Let us consider the extended Hilbert space that consists of the input system and the output system:
$ \Hsp_{\mathrm{in}}\otimes \Hsp_{\mathrm{out}}\cong \Hsp^{\otimes 2}$.
A unitary operator $ U $ acting on $ \Hsp $ is mapped to the following state of  $ \Hsp^{\otimes 2} $:
\begin{align}
\ket{U}:=
\frac{1}{d^{1/2}}\sum_{i,j}U_{ji}\ket{i}_{\mathrm{in}}\otimes\ket{j}_{\mathrm{out}}.
\end{align}
When $ U $ is the identity, this state corresponds to a maximally entangled state ($ N $ EPR pairs) between the input and output states.
We graphically show the channel-state duality in Fig.~\ref{fig:channel-state_duality}.

The $ k $-OSEE of $ U $ is defined as the $ k $-REE of $ \ket{U} $~\cite{Zanardi2001,Prosen2007}.
We divide the input (output) system into $ X $ and $ X^\mathrm{c} $ ($ Y $ and $ Y^\mathrm{c} $).
We denote the reduced state of $ \ket{U} $ on $ X $ and $ Y $ by $ \rho^{U}_{XY} $.
The $ k $-OSEE is defined as
\begin{align}
R_U^{(k)}(X,Y):=\frac{1}{1-k}\log\tr{( \rho^{U}_{XY} )^k},
\end{align}
which satisfies $ 0\leq R_U^{(k)}(X,Y) \leq \log d_Xd_Y $.
The $ k $-OSEE of the identity operator is zero.
We expect that the more complex unitary operator is, the larger the $ k $-OSEE is.
The $ k $-OSEE is used as a measure of information scrambling~\cite{Lloyd2018,Lloyd2018full,Dubail2017,Nakamura2019,Schnaack2018}.
In fact, the $ k $-OSEE is related to the operator-averaged $ 2k $-point OTOC~\cite{Hosur2016,Roberts2017}.

With the use of the partial cyclic permutation operator~\eqref{eq:partial_cycle}, for $ k\geq 2 $, the $ k $-OSEE is rewritten as
\begin{align}
R_U^{(k)}(X,Y)
=
\frac{1}{1-k}\log \frac{\tr{C_X^{-1}U^{\dag \otimes k}C_YU^{\otimes k}}}{d^k}.
\label{eq:k-OSEE_unitary}
\end{align}
We also define the $ k $-OSEE of an ensemble $ \nu $ of unitaries~\cite{You2018} by
\begin{align}
R_{\nu}^{(k)}(X,Y):=
\frac{1}{1-k}\log \frac{\tr{C_X^{-1}\Phi^{(k)}_{\nu}(C_Y)}}{d^k}.
\label{eq:k-OSEE_ens}
\end{align}
In general,
$ R_{\nu}^{(k)}(X,Y) $ does not equal $ \ave{\nu}{R_{U}^{(k)}(X,Y)} $.

\subsection{Page curve of the $ k $-OSEE}

Considering  the $ k $-OSEE of the HRU, we derive the Page curve of the $ k $-OSEE~\cite{Lloyd2018full,Zhou2017}.
We first consider the infinite temperature case, where the energy shell $\Hsh$ is given by the entire Hilbert space of dimension $2^N$ as in Sec.~\ref{sec:Entanglement:Infinite}.
From Eq.~\eqref{eq:Haar_k-channel}, we have
\begin{align}
&\tr{C_X^{-1}\Phi^{(k)}_{\nu}(C_Y)}
\nonumber\\
&=
\sum_{\pi ,\tau \in S_k}\Wg_{\pi, \tau}(d)
d_X^{c(\pi_{\mathrm{cyc}}^{-1}\pi)}d_{X^{\mathrm{c}}}^{c(\pi)}
d_Y^{c(\tau \pi_{\mathrm{cyc}})}d_{Y^{\mathrm{c}}}^{c(\tau)},
\end{align}
where we used $ \tr{W_{\pi}}=d^{c(\pi)} $ and $ \pi_{\mathrm{cyc}} $ is the cyclic permutation.
From Eq.~\eqref{eq:Weingarten_asymptotic} and \eqref{eq:Weingarten_identity}, we have
\begin{align}
\tr{C_X^{-1}\Phi^{(k)}_{\nu}(C_Y)}
\approx 
\sum_{\pi \in S_k}
d^{-k+2c(\pi)}
(d_Xd_Y)^{c(\pi\pi_{\mathrm{cyc}})-c(\pi)}.
\label{eq:HRU_k-channel_cycle}
\end{align}
The leading term of this is given by $ \pi=I $ and equals $ d^kd_X^{1-k}d_Y^{1-k} $.
Dividing Eq.~\eqref{eq:HRU_k-channel_cycle} by $ d^k $ and taking the logarithm, we obtain
\begin{align}
R_{\haar}^{(k)}(X,Y)
=
(m+n)\log 2 -\gamma_{X,Y}^{(k)},
\label{eq:k-OSEE_Page_curve}
\end{align}
where we denoted the number of qubits in $ X $ and $ Y $ by $ m $ and $ n $, respectively, and
\begin{align}
&\gamma_{X,Y}^{(k)}
\nonumber\\
&:=\frac{1}{k-1}\log \left(
\sum_{\pi \in S_k}
d^{-2k+2c(\pi)}
(d_Xd_Y)^{c(\pi\pi_{\mathrm{cyc}})-c(\pi)+k-1}
\right).
\label{eq:k-OSEE_Page_correction}
\end{align}
Equation~\eqref{eq:k-OSEE_Page_curve} is the counterpart of the Page curve~\eqref{eq:Page_curve} for the $ k $-OSEE.
The first term on the right-hand side of Eq.~\eqref{eq:k-OSEE_Page_curve}
is proportional to the sum of the size of regions $ X $ and $ Y $,
implying the volume law.
$ \gamma_{X,Y}^{(k)} $ is the subleading correction to the volume law,
i.e., the Page correction.
Especially, for the half partitions ($ m=n=N/2 $), we have $ \gamma_{X,Y}^{(k)}\to \log C_k $ in the limit of $ d\to\infty $, where $ C_k:=(2k)!/k!(k+1)! $ is the Catalan number~\cite{Lloyd2018full}.
For $ k=2 $, Eq.~\eqref{eq:k-OSEE_Page_curve} and Eq.~\eqref{eq:k-OSEE_Page_correction} reduce to
\begin{align}
R_{\haar}^{(2)}(X,Y)&=(m+n)\log 2-\gamma_{X,Y}^{(2)},
\label{eq:2-OSEE_Page_curve}
\\
\gamma^{(2)}_{X,Y}&=\log\left(
1+\frac{1-2^{2m}-2^{2n}+2^{2(m+n)}}{2^{2N}-1}
\right).
\end{align}
$ \gamma_{X,Y}^{(2)} $ is approximately zero for $m+n\ll N$,
whereas $ \gamma_{X,Y}^{(2)}\approx \log 2 $ for $ m+n=N $.
These properties are similar to the case of  the Page curve of the 2-REE~\eqref{eq:Page_curve}.

We next generalize the above argument to the finite temperature case, where the energy shell $ \Hsh $ is a strict subspace of the entire Hilbert space  as in Sec.~\ref{sec:Entanglement:Thermal}.
For simplicity, we consider the case of $ k=2 $.
The 2-OSEE of the HRU on $ \Hsh $ is defined as
\begin{align}
R_{\haar}^{(2)}(X,Y)
:=-\log\frac{\tr{\tilde{S}_X\Phi_{\haar}^{(2)}(\tilde{S}_Y)}}{\Dsh^2},
\end{align}
where $ \tilde{S}_X $ and $ \tilde{S}_Y $ are the partial swap operators projected onto $ \Hsh $.
By using Eqs.~\eqref{eq:ppswap_trace_1} and \eqref{eq:ppswap_trace_2},
we have
\begin{align}
&\Dsh^{-2}\tr{\tilde{S}_X\Phi_{\haar}^{(2)}(\tilde{S}_Y)}
\nonumber\\
&\approx
e^{-R_{\mathrm{mc}}^{(2)}(X)-R_{\mathrm{mc}}^{(2)}(Y)}
+e^{-R_{\mathrm{mc}}^{(2)}(X^\mathrm{c})-R_{\mathrm{mc}}^{(2)}(Y^\mathrm{c})}
\end{align}
for $ \Dsh\gg 1 $.
Taking the logarithm of the above equality,
we obtain
\begin{align}
R_{\haar}^{(2)}(X,Y)
&\approx R_{\mathrm{mc}}^{(2)}(X)+R_{\mathrm{mc}}^{(2)}(Y)
-\tilde{\gamma}_{X,Y}^{(2)},
\label{eq:OSEE_thermal_Page_curve}
\end{align}
where
\begin{align}
\tilde{\gamma}_{X,Y}^{(2)}&:=
\log\left(
1+e^{R_{\mathrm{mc}}^{(2)}(X)+R_{\mathrm{mc}}^{(2)}(Y)-R_{\mathrm{mc}}^{(2)}(X^\mathrm{c})-R_{\mathrm{mc}}^{(2)}(Y^\mathrm{c})}
\right).
\end{align}
Equation~\eqref{eq:OSEE_thermal_Page_curve} is  a finite-temperature generalization of the 2-OSEE Page curve~\eqref{eq:2-OSEE_Page_curve}.
The first and the second terms on the right-hand side of Eq.~\eqref{eq:OSEE_thermal_Page_curve} represent the thermal behavior,
and the third term $ \tilde{\gamma}_{X,Y}^{(2)}$ represents the Page correction, which
becomes $ \log 2 $ at $ m=n=N/2 $.
In the same manner as in Sec.~\ref{sec:Entanglement:Thermal},
we can show that
$ R_{\haar}^{(2)}(X,Y) $ follows Eq.~\eqref{eq:universal_form} by replacing $ n $ with $ n+m $ under assumption~\eqref{eq:volume-law}.
The foregoing argument can also apply to the higher $ k $-OSEE ($ k\geq 3 $), and we can derive the finite-temperature Page curve~\eqref{eq:thermal_k-REE} for the $ k $-OSEE.

\subsection{$ k $-OSEE and $ k $-ETH}
We next consider the $ k $-OSEE of the LTE.
If the system satisfies the exact $ k $-ETH for $ \tilde{C}_Y $, the $ k $-OSEE of the LTE follows the finite-temperature Page curve of the $ k $-OSEE.
This is because the exact $ k $-ETH for $ \tilde{C}_Y $ implies that $ \Phi_{\LTE}^{(k)}(\tilde{C}_Y)=\Phi_{\haar}^{(k)}(\tilde{C}_Y) $.
Since the 2-ETH is true for the nonintegrable model discussed in Sec.~\ref{sec:Numeric}, the 2-OSEE of the LTE follows the Page curve in this model.

In Refs.~\cite{Zhou2017,Schnaack2018}, it has been numerically shown that the 1-OSEE $ R_U^{(1)}(X,Y) $ of  $ e^{-iHt} $ saturates at almost the maximum value in nonintegrable systems.
This can also be accounted for by the 2-ETH as follows. 
At late times, we expect that $ R_U^{(1)}(X,Y)\approx  R_{\LTE}^{(1)}(X,Y)$ holds
if the initial state is given by a large number of energy eigenstates~\cite{Reimann2008,Linden2009}.
By combining this and $ R_\LTE^{(1)}(X,Y)\geq R_\LTE^{(2)}(X,Y) $,
we obtain
\begin{align}
R_U^{(1)}(X,Y)\gtrsim R_{\LTE}^{(2)}(X,Y)\approx R_{\haar}^{(2)}(X,Y),
\label{eq:OSEE_ineq}
\end{align}
where we used the 2-ETH to obtain the right equality.
Since $R_\haar^{(2)} (X,Y)$ gives the Page curve that is also an upper bound of $R_U^{(1)} (X,Y)$, Eq.~\eqref{eq:OSEE_ineq} implies that  $ R_U^{(1)}(X,Y) $ takes almost the maximum value.
On the other hand, the 1-OSEE does not saturate at the maximum value in integrable systems~\cite{Zhou2017,Schnaack2018}, as consistent with our numerical result  that the 2-ETH does not hold in the integrable system.

We note that the dominant contribution to the 2-OSEE comes from Eqs.~\eqref{eq:2-ETH_2_swap_thermal} and \eqref{eq:2-ETH_3_swap_thermal} at finite temperature (Eqs.~\eqref{eq:2-ETH_2_swap} and \eqref{eq:2-ETH_3_swap} at infinite temperature), which are not related to the 2-REE.
From Eq.~\eqref{eq:LTE_k-channel}, we have
\begin{align}
&\Dsh^{-2}\tr{\tilde{S}_X\Phi_{\LTE}^{(2)}(\tilde{S}_Y)}
\nonumber\\
&=
\Dsh^{-2}\sum_i\braket{E_iE_i|\tilde{S}_X|E_iE_i}\braket{E_iE_i|\tilde{S}_Y|E_iE_i}
\nonumber\\
&\quad
+\Dsh^{-2}\sum_{i\neq j}\braket{E_iE_j|\tilde{S}_X|E_iE_j}\braket{E_iE_j|\tilde{S}_Y|E_jE_i}
\nonumber\\
&\quad
+\Dsh^{-2}\sum_{i\neq j}\braket{E_iE_j|\tilde{S}_X|E_jE_i}\braket{E_iE_j|\tilde{S}_Y|E_iE_j}.
\end{align}
The three terms on the right-hand are evaluated by Eqs.~\eqref{eq:2-ETH_1_swap_thermal}, \eqref{eq:2-ETH_2_swap_thermal}, and \eqref{eq:2-ETH_3_swap_thermal}, respectively.
For $ d\gg 1 $,
we have
\begin{align}
\text{1st term}
&\approx 0,
\\
\text{2nd term}
&\approx
\begin{dcases}
e^{-R_{\mathrm{mc}}^{(2)}(X)-R_{\mathrm{mc}}^{(2)}(Y)} & (m,n \ll N) \\
e^{-2R_{\mathrm{mc}}^{(2)}(X)}   & (m=n=N/2),
\end{dcases}
\\
\text{3rd term}
&\approx
\begin{dcases}
0 &(m,n \ll N) \\
e^{-2R_{\mathrm{mc}}^{(2)}(X)}  & (m=n=N/2).
\end{dcases}
\end{align}
Here, if $ m, n\ll N$,
only the second term contributes to $R_{\LTE}^{(2)}(X,Y)  $, implying the thermal behavior $ R_{\LTE}^{(2)}(X,Y)\approx R_{\mathrm{mc}}^{(2)}(X)+R_{\mathrm{mc}}^{(2)}(Y) $.
On the other hand, if $ m=n=N/2 $, the second and the third terms contribute in the same order as $ e^{-2R_{\mathrm{mc}}^{(2)}(X)}  $, leading to both the thermal behavior and the Page correction
$  R_{\LTE}^{(2)}(X,Y)\approx -\log(e^{-2R_{\mathrm{mc}}^{(2)}(X)} +e^{-2R_{\mathrm{mc}}^{(2)}(X)} )
\approx 2R_{\mathrm{mc}}^{(2)}(X)-\log 2 $.
The reason why the first term is negligible is that the sum in that term runs over $ d $ elements while the sums in the other terms runs over $ d^2-d $ elements.

Now, the implications of the 2-ETH for $ \tilde{S}_X $ is summarized as follows:
Eq.~\eqref{eq:2-ETH_1_swap_thermal} implies the Page curve of the 2-REE~\eqref{eq:thermal_Page_curve},
Eq.~\eqref{eq:2-ETH_2_swap_thermal} implies the thermal behavior of the 2-OSEE,
and Eq.~\eqref{eq:2-ETH_3_swap_thermal} implies the Page correction of the 2-OSEE.

The first and the second terms on the right-hand side of Eq.~\eqref{eq:OSEE_thermal_Page_curve}  (the volume law at infinite temperature) are also accounted for by the 1-ETH as follows.
The 1-ETH for the partial swap operator $ \tilde{S}_X $ is regarded as quantum ergodicity for  $ \tilde{S}_X $:
\begin{align}
\Phi_{\LTE}^{(2)}(\tilde{S}_X)
=\frac{\tr{\tilde{S}_X}}{\Dsh^2}I^{\otimes 2}
=e^{-R_{\mathrm{mc}}^{(2)}(X)} I^{\otimes 2}.
\end{align}
In this case, we have
\begin{align}
R_{\LTE}^{(2)}(X,Y)=R_{\mathrm{mc}}^{(2)}(X) +R_{\mathrm{mc}}^{(2)}(Y),
\end{align}
which equals the sum of the first and the second terms on the right-hand side of Eq.~\eqref{eq:2-OSEE_Page_curve}.
Therefore, the Page correction of the OSEE originates from the higher-order ETH.

\subsection{Operator-averaged OTOC}

We remark on the relationship between the OSEE and the operator-averaged OTOC.
For the infinite temperature case with the Hilbert space of dimension $ d:=2^N $,
the operator-averaged (4-point) OTOC on region $ X $ and $ Y^{\mathrm{c}} $ is defined by
\begin{align}
F_{X,Y}(t):=\frac{1}{d_X^2 d^2_{Y^{\mathrm{c}}}}
	\sum_{P_X,P_{Y^{\mathrm{c}}}}\braket{P_X(t)P_{Y^{\mathrm{c}}}P_X(t)P_{Y^{\mathrm{c}}}}_{\mathrm{mc}},
\end{align}
where the sum is taken over all the Pauli operators $ P_X $ on $ X $ and $ P_{Y^{\mathrm{c}}}$  on $ Y^{\mathrm{c}} $.
As mentioned before, the operator-averaged OTOC is related to the 2-OSEE~\cite{Hosur2016,Roberts2017} as
\begin{align}
\ave{\nu}{F_{X,Y}}=2^{-m-n}e^{-R^{(2)}_\nu(X,Y)}
\end{align}
with an arbitrary ensemble $ \nu $.
Thus, the long-time average $ \overline{F}_{X,Y}$ exactly decays to the HRU average $ F_{X,Y}^{\haar} $, if the 2-OSEE of the LTE exactly follows the Page curve: $R_{\LTE}^{(2)}(X,Y)=R_{\haar}^{(2)}(X,Y)  $, (or equivalently, the exact 2-ETH holds for $ S_{Y} $).

We next consider approximate decay of the operator-averaged OTOC.
For $ d\gg 1 $, the HRU average of the operator-averaged OTOC is approximated as
\begin{align}
F_{X,Y}^{\haar}
&\approx
\begin{dcases}
2^{-2(m+n)} & (m,n\ll N)\\
2\cdot 2^{-2N} & (m=n=N/2),
\end{dcases}
\end{align}
where the prefactor 2 for $ m=n=N/2 $ originates from the Page correction.
If $ X $ and $ Y $ are local ($ m,n \ll N $), the approximate decay of the operator-averaged OTOC is guaranteed by the volume law of the 2-OSEE, $ R_{\LTE}^{(2)}(X,Y)\approx (m+n)\log 2 $, as a consequence of the 1-ETH .
On the other hand, for $ m=n=N/2 $, the 1-ETH does not give the prefactor 2, which is a consequence of the 2-ETH.

\section{Supplemental numerical results
\label{sec:Supp_numeric}}
In this Appendix, we provide supplemental numerical results.

\subsection{4-point OTOC}
We here provide supplemental  numerical results for the (4-point) OTOC
\begin{align}
F_{A,B}(t):=
\braket{A(t)BA(t)B}_{\mathrm{mc}},
\label{eq:OTOC}
\end{align}
in the XXZ ladder model~\eqref{eq:XXZ-ladder} by using numerical exact  diagonalization.
To simplify notation, we omit the superscript `$ (4) $'.

\begin{figure}
	\includegraphics[width=1.0\linewidth]{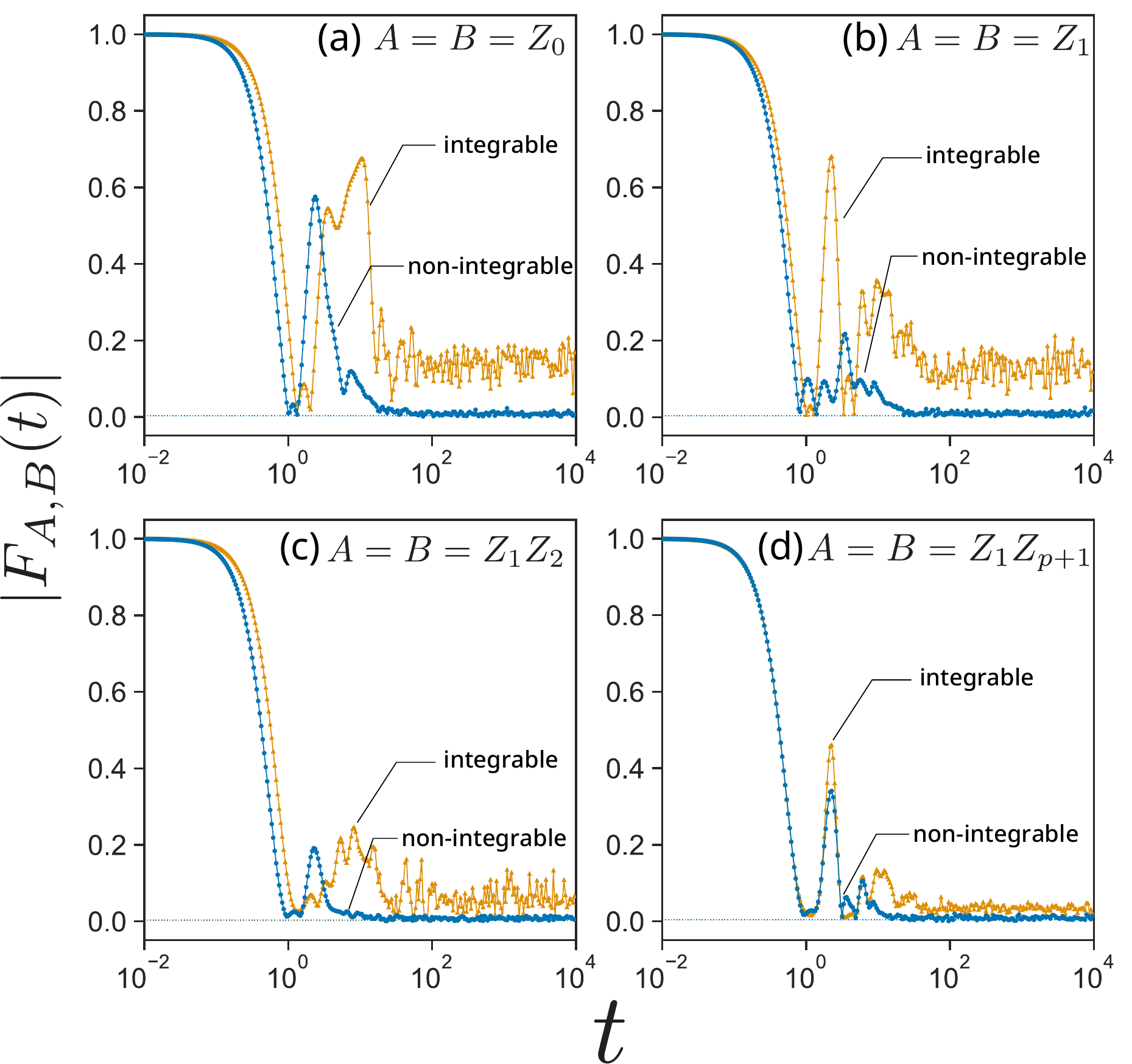}
	\caption{\label{fig:OTOC_XXZ}
		The time-dependence of the OTOC~\eqref{eq:OTOC} for
		the XXZ ladder model \eqref{eq:XXZ-ladder} with
		$ \lambda=0 $ (integrable) and	$ \lambda=1 $ (nonintegrable).
		The dotted lines show the HRU average.
		We set 	$ A=B= $
		(a) $ Z_0 $,
		(b) $ Z_1 $,
		(c) $ Z_1Z_2 $,
		and (d) $ Z_1Z_{p+1} $.
		The OTOC  relaxes (does not relax) to the HRU average
		for the nonintegrable (integrable) case.
	}
\end{figure}

Figure~\ref{fig:OTOC_XXZ} shows the time-dependence of the OTOC~\eqref{eq:OTOC}.
We set $ A=B=Z_0, Z_1, Z_1Z_2, Z_1Z_{p+1} $, and the parameters are the same as those in Sec.~\ref{sec:Numeric}.
For the nonintegrable case ($ \lambda=1 $), $ F_{A,B}(t) $ relaxes to a small value near the HRU average.
On the other hand, for the integrable case ($ \lambda=0 $),
$ F_{A,B}(t) $ exhibits larger temporal fluctuations and does not relax to the HRU average.
These results again confirm that information scrambling occurs only in nonintegrable systems.

In order to further clarify the relationship between the integrability and  the OTOC at late times, we investigate the $ \Dsh $-dependence of the deviation of the long-time averaged OTOC~\eqref{eq:OTOC_LTA}
 from the HRU average~\eqref{eq:OTOC_HRU}:
\begin{align}
\Delta F:=\abs{\overline{F}_{A,B}-F_{A,B}^{\haar}}.
\label{eq:OTOC_error}
\end{align}
We calculated $ \bar{F}_{A,B} $ with the help of the analytical form~\eqref{eq:2k-OTOC_LTE}.
We note that we numerically confirmed the 2nd-incommensurate condition of the spectrum of the XXZ ladder model.

\begin{figure}
	\includegraphics[width=1.0\linewidth]{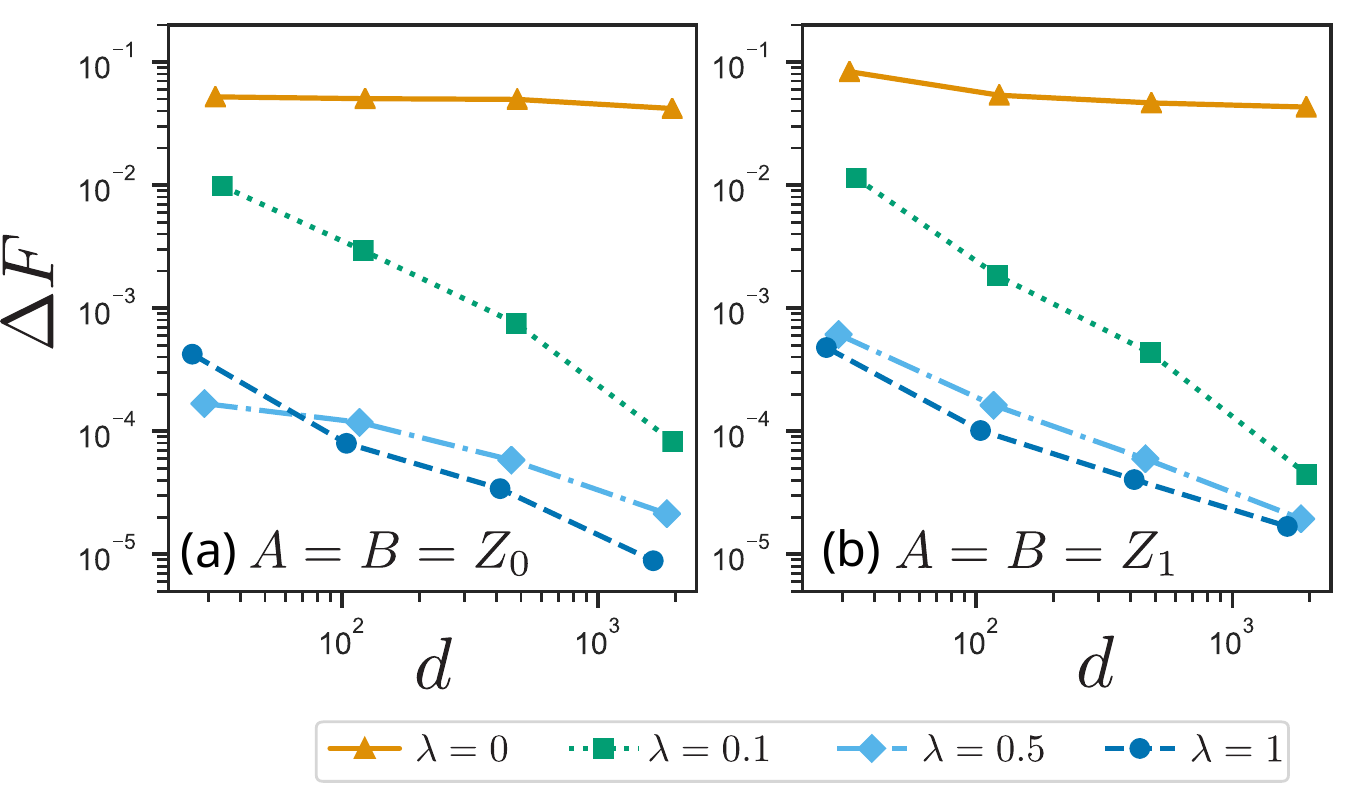}
	\caption{\label{fig:OTOC_scaling_XXZ}
		The $ \Dsh $-dependence of $ \Delta F $ defined in \eqref{eq:OTOC_error} for the XXZ ladder model \eqref{eq:XXZ-ladder} with $ \lambda=0,0.1,0.5,1 $.
		We set 	$ A=B$ to (a) $ Z_0 $, (b) $ Z_1 $.
		These results imply that the deviation from the HRU average
		vanishes (does not vanish) with $ \Dsh\to\infty $
		for the nonintegrable (integrable) case.
	}
\end{figure}

Figure~\ref{fig:OTOC_scaling_XXZ} shows the $ \Dsh $-dependence of $ \Delta F $ with the integrability breaking parameter $ \lambda =0, 0.1, 0.5, 1 $.
For the nonintegrable case, $ \Delta F $ decreases polynomially with $ \Dsh $, implying that the long-time average of the OTOC agrees with the HRU average in the limit of  $ \Dsh\to\infty $.
On the other hand, $ \Delta F $ does not decay for the integrable case.
We note that $ \Delta F $ is larger than $ \mathcal{O}(\Dsh^{-2}) $ for all the parameters.
This fact implies that,  for finite $ \Dsh $,  the OTOC does not achieve the exact decay to the HRU average even in the nonintegrable case.

These results are consistent with our theory in the main text as follows.
The conventional diagonal and off-diagonal ETH is only true for the nonintegrable cases ($ \lambda\neq 0 $).
Thus, Eq.~\eqref{eq:approximate_scrambling}, which implies the decay of the long-time average of the OTOC, is only applicable to the nonintegrable cases.
For the integrable case, Eq.~\eqref{eq:approximate_scrambling} does not give any information about the OTOC.

\subsection{The $ \Dsh $-dependence of the microcanonical average
\label{sec:mc_exponent}}
We show the $ \Dsh $-dependence of the microcanonical average $ \braket{A}_{\mathrm{mc}} $ to estimate the exponent $ a $ of $  I_2(A^{\otimes 2}) = \mathcal{O}(\Dsh^{-a})$  from Eq.~\eqref{eq:scaling_RMT_product}.

\begin{figure}
	\includegraphics[width=0.6\linewidth]{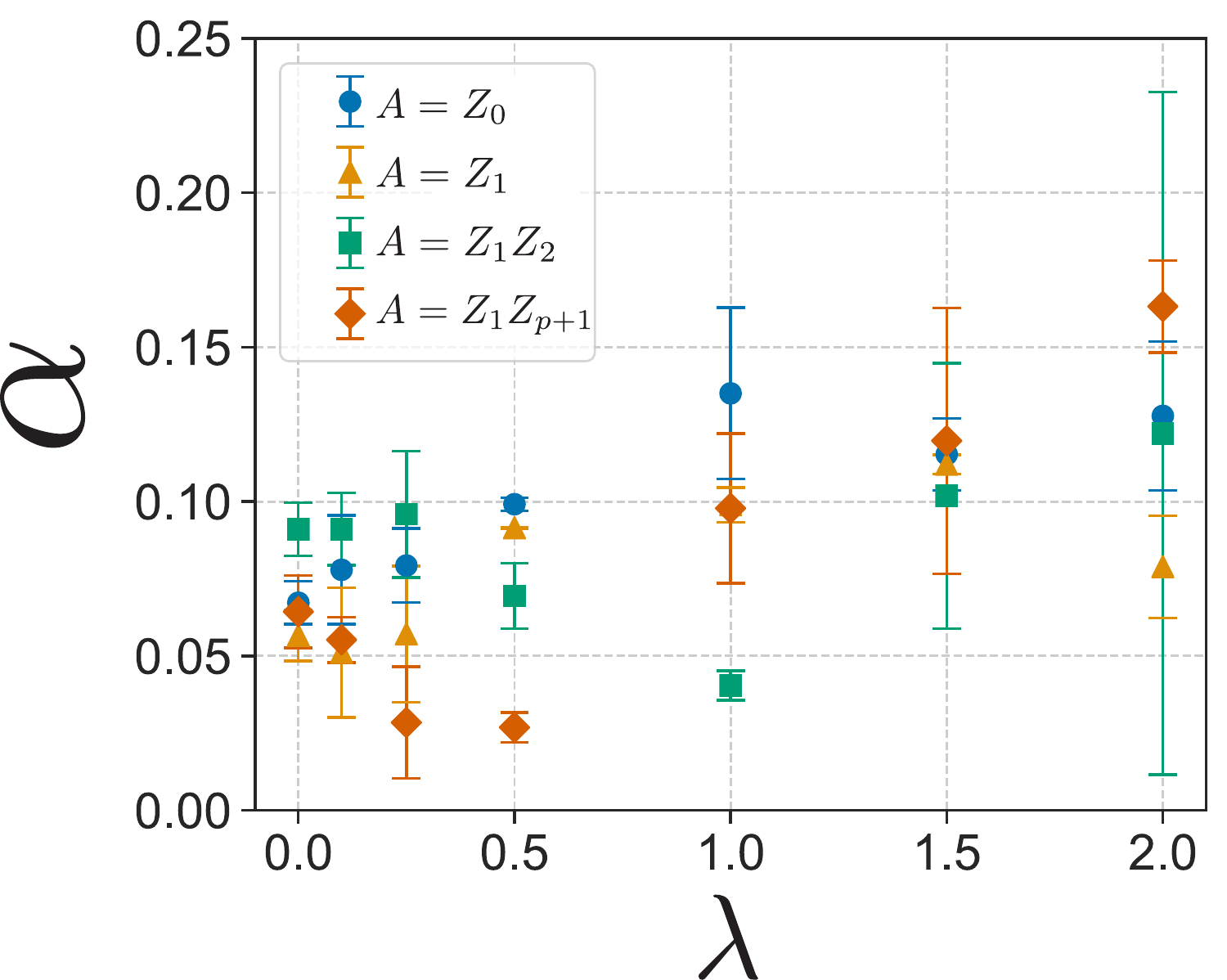}
	\caption{\label{fig:mc_exponent}
		The $\lambda$-dependence of the exponent $ \alpha $ of $  \abs{\braket{A}_{\mathrm{mc}}}=\mathcal{O}(\Dsh^{-\alpha})  $.
		The exponent takes $ \alpha \sim 0.1 $ in the nonintegrable case ($ \lambda \neq 1 $).
		As in Fig.~\ref{fig:tensor}, the circles, triangles, squares, and diamonds represent the results  for  $ Z_0 $, $ Z_1$, $ Z_1Z_2$, and $Z_1Z_{p+1} $, respectively.
	}
\end{figure}

As in Sec.~\ref{sec:Numeric}, we calculate the microcanonical average of $ A $ corresponding to the the energy shell $ [E_i-\delta_{\mathrm{mc}} , E_i] $ by changing an eigenenergy $E_i$.
We choose the maximum absolute value $ \abs{\braket{A}_{\mathrm{mc}}} $  over  $ E_i \in [E-\Delta E, E]  $, and fit $ \log \abs{\braket{A}_{\mathrm{mc}}} $ against a fitting function 
$f(\Dsh):=-\alpha\log \Dsh +\beta$ with fitting parameters $\alpha $ and $\beta $.
We use the same parameters as in Sec.~\ref{sec:Numeric}.

Figure~\ref{fig:mc_exponent} shows the $\lambda$-dependence of the exponent $ \alpha $ of $ \abs{\braket{A}_{\mathrm{mc}}} =\mathcal{O}(\Dsh^{-\alpha})  $.
We find that the exponent $ \alpha $ is positive for all the cases, implying that $ \abs{\braket{A}_{\mathrm{mc}}} \to 0 $ in the limit of $ \Dsh\to 0 $.
This is consistent with the expectation from the equivalence of ensembles~\cite{Ruelle1999,Tasaki2018}.
 $ \alpha $ depends on parameter $ \lambda $ and operator $A$, but takes a value around $ 0.1 $.
We thus used $ \alpha\sim 0.1 $ to estimate the exponent $  a $ in Sec.~\ref{sec:Numeric:product}.

\end{document}